\documentclass{elsart}

\usepackage{natbib}

\usepackage{amssymb}

\usepackage{graphicx}
\usepackage{amsmath}
\usepackage{epstopdf}
\usepackage{placeins}
\usepackage{array}
\graphicspath{{figs/}}

\begin{document}

\begin{frontmatter}



\title{Heat Transfer in Underground Rail Tunnels}


\author{Stefan Sadokierski}
\ead{stefan.sadokierski@arup.com}
\address{Arup, 13 Fitzroy Street London W1T 4BQ, United Kingdom}

\author{Jean-Luc Thiffeault}
\ead{jeanluc@imperial.ac.uk}
\address{Department of Mathematics, Imperial College London, London SW7 2AZ,
  United Kingdom}

\begin{abstract}
The transfer of heat between the air and surrounding soil in underground
tunnels ins investigated, as part of the analysis of environmental conditions
in underground rail systems.  Using standard turbulent modelling assumptions,
flow profiles are obtained in both open tunnels and in the annulus between a
tunnel wall and a moving train, from which the heat transfer coefficient
between the air and tunnel wall is computed.  The radial conduction of heat
through the surrounding soil resulting from changes in the temperature of air
in the tunnel are determined. An impulse change and an oscillating tunnel air
temperature are considered separately.  The correlations between fluctuations
in heat transfer coefficient and air temperature are found to increase the
mean soil temperature.  Finally, a model for the coupled evolution of the air
and surrounding soil temperature along a tunnel of finite length is given.  
\end{abstract}


\end{frontmatter}

\section{Introduction}
\label{sec:intro}

Aero-thermodynamic analysis of the environmental conditions within underground
tunnels has a number of engineering applications. Perhaps the most widespread
application involves modelling the air temperatures and conditions expected in
underground rail stations. The paper by \citet{UC} is
one example of many publications on this subject: there the air temperature
within London's King's Cross Underground Station was simulated and compared
with measurements taken within the existing station in order to validate the
model.  The model was then changed to reflect proposed modifications to the
station and re-run to simulate future conditions.  The main aim was to ensure
that overheating of the station does not occur as a result of increased
patronage and servicing.

The software used to develop such models must simulate a large number of
phenomena. Typically, an aerodynamic network model of the tunnels, stations,
passageways and ventilation shafts is the basis of the model and is used to
predict air movements generated by trains moving through the system. The heat
generated and released by trains (by far the largest source of heat energy
within an underground rail system) is typically also computed. The effect of
other heat sources such as electro-mechanical equipment, lighting and
passengers is accounted for. 

The final factor, which cannot be accounted for by a steady-state heat balance
approach, is the thermal interaction between air within the tunnel and the
surrounding soil. The tunnel lining and surrounding soil have a large thermal
mass and are strongly coupled to the air within the system, so they must be
included if conditions in the underground environment are to be accurately
modelled. However, as we will show, the effect of the surrounding soil depends
on a large number of parameters, including the magnitude and frequency of
fluctuations in ambient air temperature due to daily and seasonal
oscillations.

Details of a comprehensive thermal model of an underground railway environment
are given by \citet{Ampofo}. Results presented in their paper give an overview
of the relative importance of heat loads and sinks within an underground
system. For instance, during peak summer conditions 85\% of the heat load in a
tunnel is generated by train braking systems, 13\% by other systems on the
trains and 2\% from tunnel lighting and other loads within the tunnels. Hence,
the introduction of regenerative braking systems has the greatest potential
to reduce the total thermal load and hence peak temperatures. The model also
predicts that 70\% of the heat load is removed by air moving through the
tunnels and the remaining 30\% is lost by conduction into the tunnel
walls. However, the thermal model treats the tunnel wall as a thermal
resistance only, and hence does not take into account the thermal mass of the
tunnel walls and surrounding soil. As a result the daily and seasonal phase
shift in tunnel air temperatures observed by \citet{UC}---the fact that the
Underground warms up over the summer and is therefore warmer in autumn than
spring for the same ambient conditions---cannot be predicted by ther
model. From a more practical point of view, this limitation also means that
standard steady-state modelling cannot predict what effect `night
flushing'---using fans to move cool night air through the tunnels when trains
are not operating---will have on the peak temperature in the tunnels the next
day. It is these effects that we address in this paper.

Our analysis encompasses the following effects:

\begin{itemize}

\item \emph{Section \ref{sec:turbulentflow2} -- Turbulent flow profiles}.  We
  present analytic formulations of the turbulent flow profiles expected in
  both an open tunnel and in the annulus between the tunnel and a passing
  train are presented. These are used to compute the coefficient governing
  heat transfer between the air and the internal surface of the tunnel.

\item \emph{Sections \ref{sec:conduction1} and \ref{sec:conduction2} -- Radial
  heat conduction}. Using the flow profiles above, we derive analytic
  solutions governing the radial conduction of heat in a cylindrical coodinate
  system. The conducting domain extends from an internal radius to
  infinity. Heat transfer into the solid from the internal radius is
  conductive and proportional to the difference between the temperature of the
  air within the tunnel and the temperature at the wall surface.

\item \emph{Section \ref{sec:fluctuations} -- Short-term fluctuations}. We
  consider the effect of short-term fluctuations in heat transfer coefficient
  caused by the passage of trains. The approach currently used in engineering
  calculations involves time averaging the heat transfer coefficient over the
  short-term fluctuations. A more comprehensive approach, not currently used,
  is presented, involving the correlation between short-term fluctuations in
  heat transfer coefficient and tunnel air temperature, both caused by passing
  trains.  We show that such a correlation manifests itself as an additional
  heat source within the tunnel and hence raises the mean temperature.

\item \emph{Section \ref{sec:tunnel} -- Coupled evolution of air and soil
  temperatures}. Finally, we present a model for the coupled evolution of air
  temperature (along the axis of a tunnel) and soil (radially from the tunnel
  surface).  Unlike previous models, such as \citet{peavy} this accounts for a continuous inhomogeneous heat source along
  the tunnel axis.

\end{itemize}

To give the reader a feel for the magnitude of the effects involved, we use
specific values for physical parameteres throughout the paper.  Unless
otherwise stated, we use parameters typical of London's Piccadilly line tunnels
and trains.  For easy reference, these parameter values are documented in
tables in Appendix~\ref{apxA}.

\section{Heat Transfer Coefficient at the Tunnel Wall}
\label{sec:turbulentflow2}

In its most fundamental form, the problem under consideration is described by
Newton's law of cooling: the heat flux between the air within a tunnel and the
surrounding soil is proportional to the difference between the temperature of
the air and of the tunnel wall surface.  The magnitude of the heat transfer or
flux for a given temperature difference is dictated by the heat transfer
coefficient $h$, a parameter that is a function of the nature of the air flow
and properties of the wall surface.

In this section we present mathematical approximations of steady state (time
independent) air flow through a train tunnel.  As per previous work on this
subject by \citet{barrow} and \citet{shigechi}, we consider two distinct
cases. Firstly, air flow through a circular tunnel, and secondly, air flow in
the annulus between a stationary circular tunnel and a concentric circular
core moving in the axial direction.  These two cases represent, respectively,
air flow through an open tunnel and in the annulus between a tunnel wall and
moving train.  In both cases we show that the flows are fully turbulent under
the generic conditions considered, and give generic turbulent flow profiles.

We then use the so-called `Reynolds analogy' to relate the air flow velocity
profile, as described above, to the air flow temperature profile.  This
analogy gives an approximation to the heat transfer coefficient $h$ as a
function of the flow shear stress at the tunnel wall $\tau_w$.  As a result,
the heat transfer coefficient can be estimated from the parameters of the air
flows through the tunnel and through the annulus between the tunnel wall and a
passing train.

\subsection{Turbulent Flow Through an Unobstructed Tunnel}

As mentioned above, the tunnel air flow profile, and hence the heat transfer
coefficient, is drastically different depending on its regime, which can be
laminar, turbulent, or intermediate between the two.  The regime is determined
using the dimensionless Reynolds number.  For flow in a circular pipe, or in
this case tunnel, the Reynolds number is defined by
\citet[Eq. 17.129]{schlichting} as~$Re = {U_t a}/{\nu}$.  Here, in cylindrical
polar coordinates, $U_t$ is the mean fluid velocity along the axis of the pipe
 for a flow $u(r)$, $a$ is the radius of the pipe wall, and
$\nu$ is the kinematic viscosity of the fluid.

In general, flow with this geometry is fully turbulent if $Re > 2000$, as can
be shown on a Moody diagram \citep[Figure 6.8]{kreith}.  For a typical air flow
through a tunnel induced by train piston effect, $Re \approx 10^6$---the flow
is assumed to be fully turbulent. This means that the flow will not
be uni-directional, and hence the well-known Poiseuille solution where the
radial velocity distribution is parabolic does not apply.  We must resort to
turbulent modeling of the mean flow profile.  Here we will use Prandtl's
mixing length hypothesis, which gives a mean velocity distribution as a
logarithmic profile from the tunnel wall, and symmetrical about the centre of
the tunnel ($r=0$).  This is discussed in numerous texts on turbulent flow,
for instance~\citet{pope}.

A generic mean velocity profile for fully turbulent flow in a pipe or radius
$a$ is
\begin{equation}
{u(y)}/{v_*} = 2.5 \log{\left({y}/{k}\right)} + 8.5\,,
\label{pipeflow}
\end{equation}
for $0 \leq y \leq a$ where $y = a-r$, the distance into the flow from the
wall boundary and $k$ is the roughness length of the wall surface. Further,
\begin{equation}
v_* = \sqrt{\tau/\rho}
\label{vstar}
\end{equation} 
is the friction velocity where $\tau$ is the turbulent shear stress of the
flow, a constant over the diameter of the pipe, and $\rho$ is the density of
the fluid.

It should be noted that in \eqref{pipeflow}, the constant term given as $8.5$
is in fact a parameter of the flow. The value given applies to flow over a
`rough' wall, which requires $v_* k/ \nu > 70$, which we will show is the case
for the parameters under consideration. For flow over a `smooth' wall, which
is indicated by $v_* k/ \nu <70$, the constant term becomes a function of the
roughness parameter as described fox example in \citet{grimson}.

Although the generic velocity profile takes the form of~\eqref{pipeflow}, a
specific friction velocity must be calculated for a particular flow. This will
then define the velocity profile across the pipe and hence the wall shear
stress by~\eqref{vstar}.  We are presuming the flux, produced by either the
train piston effect or a ventilation system, is known. This value will
normally be provided by either measurements or aerodynamic modelling. 

The flow rate down the tunnel, $Q$, can also be related to the mean flow
profile, $u(r)$, as
\begin{equation}
Q =  2 \pi \int_0^a{r u(r) dr}\,.
\end{equation}
When combined with~\eqref{pipeflow}, this provides the
relationship
\begin{equation}
v_* = \frac{Q}{2 \pi a^2 \left(2.375 + 1.25 \log \left(1/k \right) \right)}
\label{vstarvalue}
\end{equation}
for the friction velocity.

\begin{figure}[h!]
\begin{center}
\includegraphics[scale=0.5]{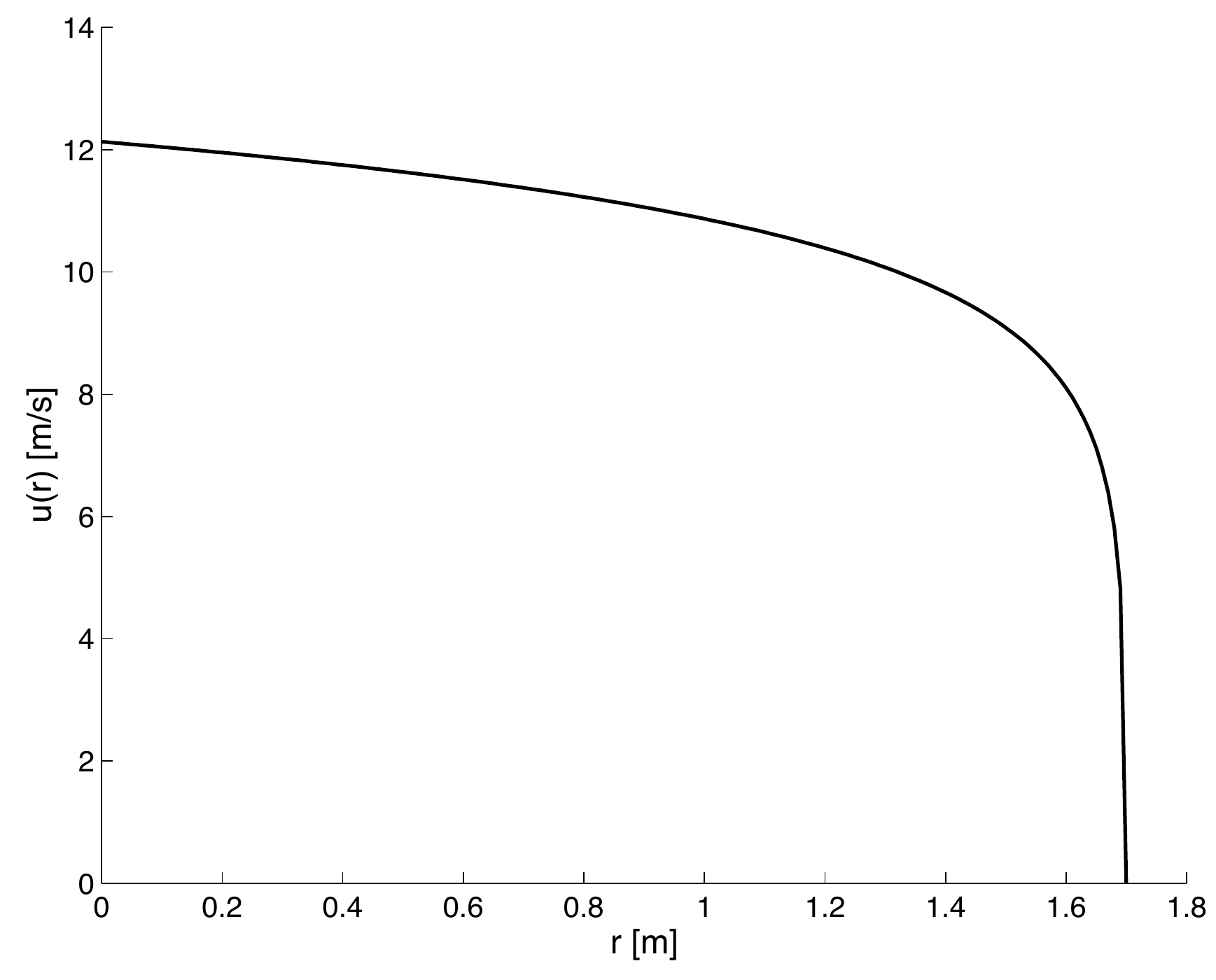}
\caption{Tunnel flow profile using the Piccadilly line parameters.}
\label{graph_tunnelflow}
\end{center}
\end{figure}
We have used parameters for a typical tunnel section of the Piccadilly line to
generate a flow profile using~\eqref{pipeflow} and~\eqref{vstarvalue}.  The
fully developed mean flow profile, $\bar{u}(r)$, is shown in
Figure~\ref{graph_tunnelflow}. The values assumed for the various parameters
and related calculations are presented in Appendix A.  The Reynolds number of
the flow is approximately $10^6$.  The shear stress at the tunnel wall for
flow in the annulus, $\tau_w$ is approximately $0.44$ N/m$^2$.  The flow
number, $v_* k / \nu$, is equal to 392, so the assumption that the wall is
rough can be made.

\subsection{Turbulent Flow in the Train--Tunnel Annulus}

The velocity profile adjacent to the tunnel wall changes considerably as a
train passes, which affects the heat transfer between the air and the wall.
In this section the air flow in the gap between the outer surface of a train
and the tunnel wall is considered. This flow is approximated by treating the
train as a circular core moving concentrically through the circular tunnel as
per the approach by \citet{barrow} and \citet{shigechi}. The radius of the
central core, $b$, is then calculated by equating the frontal area of the
train and the cross section of the cylindrical core.  The Reynolds number for
flow in such an annulus is defined by \citet{shigechi} as
\begin{equation}
Re_a = \frac{2 U_a (a - b)}{\nu}.
\label{reynolds_annulus}
\end{equation}

\subsection{Generic Annulus Velocity Profile}
\begin{figure}[h!]
\begin{center}
\includegraphics[scale=0.75]{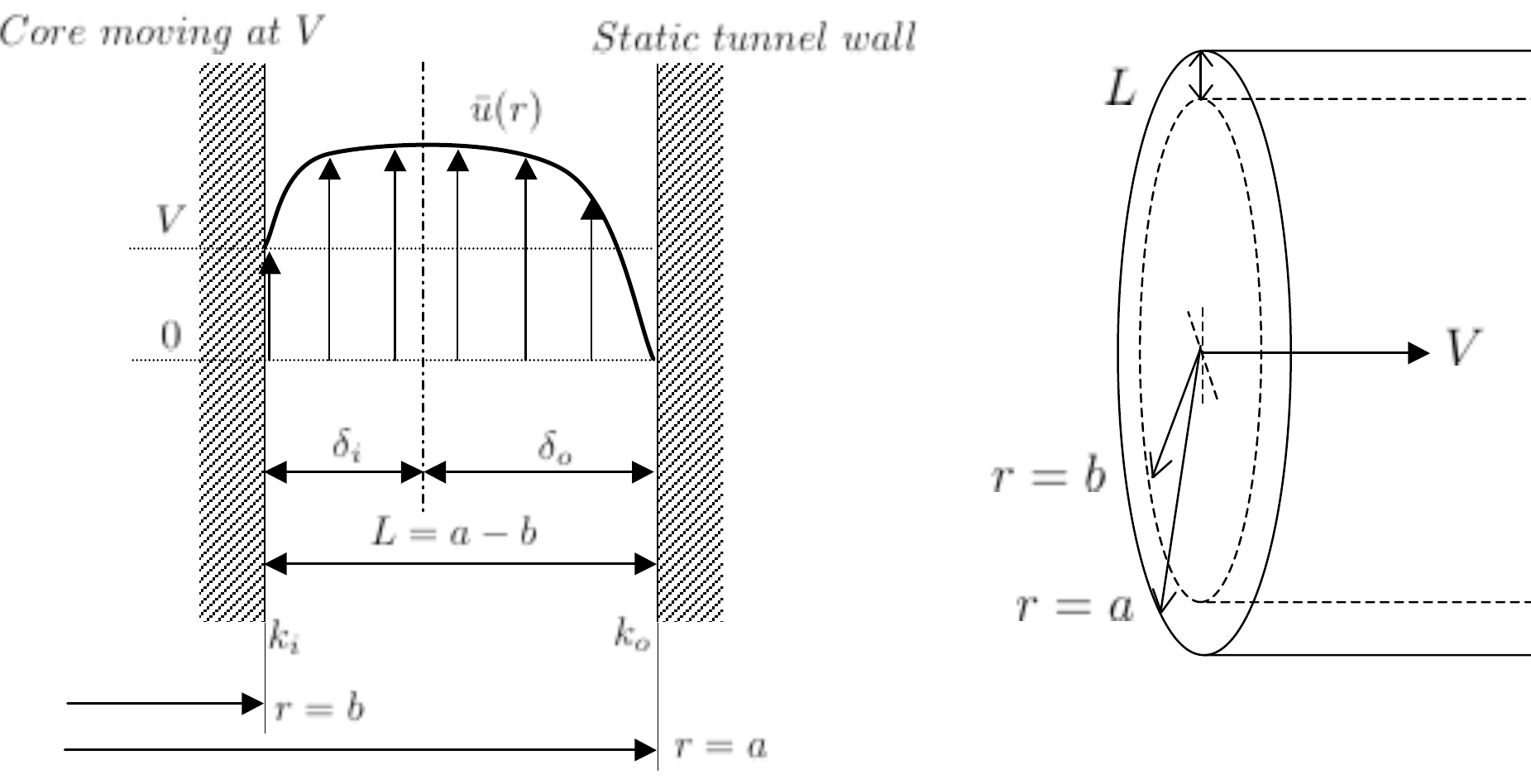}
\caption{Geometry of the fully developed annulus velocity profile.}
\label{fig_annulus}
\end{center}
\end{figure}

The general strategy for determining the flow profile through the
tunnel--train annulus, as outlined in \citet{barrow} is to match two boundary
layers of the form given by~\eqref{pipeflow}. This is shown diagramatically in
Figure~\ref{fig_annulus}.
Here the boundary on the left represents the outer surface of the train at $r
= b$, with surface roughness $k_i$ moving with velocity $V$. The boundary on
the right represents the stationary tunnel wall surface at $r = a$ ($> b$)
with surface roughness $k_o$.  The flow adjacent to the train forms a boundary
layer of thickness $\delta_i$. The no-slip boundary at the train surface is
moving with velocity $V$ relative to the tunnel wall. The flow adjacent to the
tunnel wall forms a boundary layer of thickness $\delta_o$ with the velocity
equal to zero at the wall to satisfy the no-slip boundary condition.  The
distance between the surface of the train and the tunnel wall, $L = a - b$, is
the sum of the two boundary layers, $\delta_i + \delta_o$.  The flow profile,
$u(r)$, is also shown in Figure~\ref{fig_annulus}. This profile is generated
by the matching of two separate profiles for flow over the two boundary
layers.

\begin{figure}[h!]
\begin{center}
\includegraphics[scale=0.75]{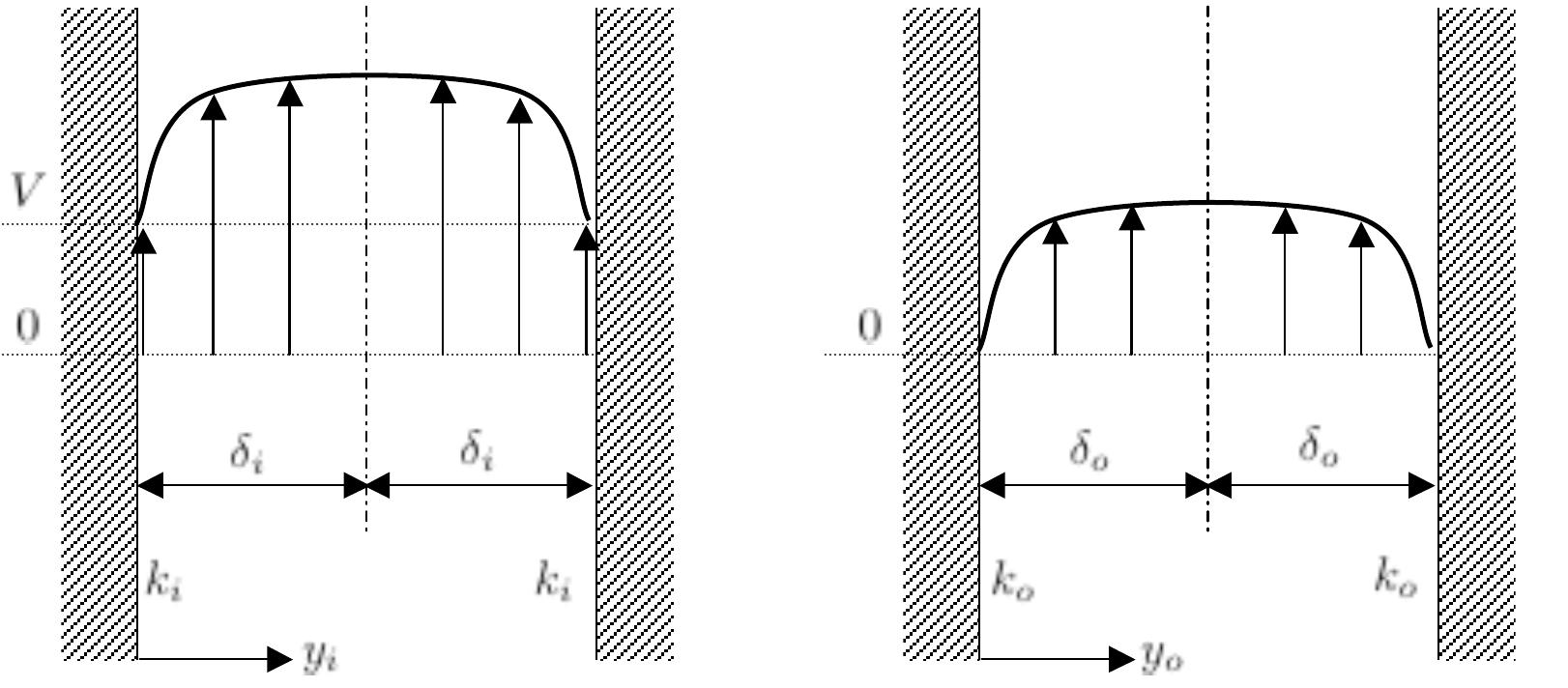}
\caption{Two ducts containing the flow adjacent to the train (on the left) and the flow
  adjacent to the tunnel wall (on the right).}
\label{fig_ducts}
\end{center}
\end{figure}

A method that can be used to arrive at the matched velocity profile is
outlined in both \citet{barrow} and \citet{shigechi}. The process involves
consideration of two `imaginary ducts,' each twice the width of the two
boundary layers $\delta_i$ and $\delta_o$ shown in
Figure~\ref{fig_annulus}. These two ducts are shown in Figure~\ref{fig_ducts}.

The flow in each of the two ducts has the generic flow profile given
by~\eqref{pipeflow}. The mean flow in the inner boundary layer adjacent to the
train takes the form
\begin{equation}
u_i(y_i) = v_{*i} \left(2.5 \log \left(\frac{y_i}{k_i}\right) +
8.5\right),
\label{innerprofile}
\end{equation}
while the flow in the outer boundary layer adjacent to the tunnel wall takes
the form
\begin{equation}
u_o(y_o) = v_{*o} \left(2.5 \log \left( \frac{y_o}{k_o}\right) + 8.5\right).
\label{outerprofile}
\end{equation}
There are four unknown parameters: the friction velocities, $v_{*i}$ and
$v_{*o}$, and the boundary layer thicknesses, $\delta_i$ and~$\delta_o$,
associated with each of the two flow profiles.  We now describe the four
equations relating the parameters required for a solution.

Firstly, the sum of the two boundary layer thicknesses is equal to the gap
between the train and tunnel wall such that
\begin{equation}
\delta_i + \delta_o = a - b.
\label{L}
\end{equation}

Secondly, the mean velocity of the inner profile must match the mean velocity
of the outer profile at the interface of the two boundary layers,
$u_i(\delta_i) + V = u_o(\delta_o)$, giving
\begin{equation}
v_{*i} \left(2.5 \log(1/ k_i) + 8.5\right) + V = v_{*o} \left(2.5 \log(1 /
k_o) + 8.5\right).
\label{matching}
\end{equation}

Thirdly, by continuity flux through the annulus must be equal to $Q$, the flow
through the open tunnel as described earlier in this section. Integrating over
the two boundary layers and allowing for the morion of the inner layer at
velocity $V$ relative to the inner boundary layer implies that
\begin{equation}
Q = 2 \pi \int_0^{\delta_i} (b+ y_i) u_i(y_i) d y_i + 2 \pi V
\int_0^{\delta_i} y_i dy_i + 2 \pi \int_{0}^{\delta_o} (a-y_o) u_o(y_o) dy_o\,.
\end{equation}
Substituting \eqref{innerprofile} and \eqref{outerprofile} into the above
expression and evaluating the integral gives
\begin{equation}
Q = 2 \pi v_{*i} A_i + \pi (\delta_i^2 + 2 b \delta_i) + 2 \pi v_{*o} A_o\,,
\label{annulusflow}
\end{equation}
where
\begin{equation}
A_i = \delta_i \left(6 b + 3.625 \delta_i + \left( \left(2.5 b + 1.25\delta_i
 \right)
 \log(1/k_i ) \right) \right) 
\label{Ai}
\end{equation}
and
\begin{equation}
A_o = \delta_o \left(6 a - 3.625 \delta_o + \left((2.5 a - 1.25\delta_o)
\log(1/k_o)\right)\right).
\label{Ao}
\end{equation}

The fourth and final relation is obtained by assuming that the flow profiles
are fully developed so that there is no mean acceleration.  Hence, the mean
forces acting on the fluid must balance to zero. This requires $\tau / \delta
= \text{constant}$, see \citet[Section 7.1.2]{pope}.  It follows that
\begin{equation}
v_{*i} = v_{*o}  \sqrt{\delta_i/\delta_o}\,.
\label{forcebalance}
\end{equation}

The four relationships given by~\eqref{L}, \eqref{matching},
\eqref{annulusflow} and \eqref{forcebalance} can be used to solve for the
previously noted unknowns numerically.  This
approach differs from that presented in \citet{barrow} and \citet{shigechi}:
both rely on an assumed value of the annulus axial pressure gradient of the
flow, which requires prior knowledge of the flow profile. The shear
stress at the train surface and tunnel wall can be computed
using~\eqref{vstar} with $v_{*i}$ and $v_{*o}$ respectively.

The parameters for a typical tunnel section of the Piccadilly line are listed
in Appendix \ref{apxA}. Details of the computations required to generate the matched
velocity profile are also given.
The thickness~$\delta_i$ of the inner boundary layer formed adjacent to the
surface of the train is approximately $80$ millimeters. The boundary layer
adjacent to the tunnel wall takes up the remaining $240$ millimeters.  The
shear stress at the tunnel wall for flow in the annulus, $\tau_w$ is
approximately $3.2$ N/m$^2$. This is roughly one order of magnitude
larger than the shear stress at the tunnel wall for the open tunnel flow given
the same air flow rate.

\subsection{Heat Transfer at the Tunnel Wall}

The motivation behind the study of open tunnel and annulus flow profiles in
the previous sections is to determine $\tau_w$, the shear stress at the tunnel
wall.  This is related to $h_w$, the coefficient of convective heat transfer
between the tunnel air and wall, by the Reynolds analogy. The main results
are briefly presented here.  A full derivation is available in most heat
transfer texts, including \citet{grimson} and \citet{kreith}. Detailed
discussion of combined thermal and velocity boundary layers is given in
\citet[Chapter 18]{schlichting}.

The rate of heat transfer between the tunnel air into the tunnel wall is
defined by Newton's law of cooling such that
\begin{equation}
\dot{q} = h \left( f\left(t \right) - T \left(a,t \right)\right),
\end{equation}
where $\dot{q}$ is the heat flux into the solid, $f(t)$ is the temperature of
the air in the tunnel averaged over the tunnel cross-section, $h$ is the
coefficient of convective heat transfer and $T(a,t)$ is the temperature of the
wall surface $(r=a)$.

\subsection{The Reynolds Analogy}

The Reynolds analogy uses the fact that, when there is both a velocity and
temperature boundary layer through a flow adjacent to a wall, there is a
similarity between the momentum and energy equations for the flow and hence
the velocity and temperature profiles.  This similarity stems from the fact
that heat transfer is proportional to the first derivative of temperature and
shear stress is proportional to the first derivative of velocity over the
boundary layer.

The Prandtl number of the fluid is defined as
\begin{equation}
Pr = \frac{C_p \mu}{K} = \frac{\nu}{\kappa}\,.
\label{prandtl number}
\end{equation}
Here $C_p$ is the specific heat capacity, $K$ is the conductivity and $\kappa$
is the thermal diffusivity of the fluid.  When~$Pr$ is equal to~$1$, the
energy and momentum equations and therefore the temperature and velocity
profiles are similar. In this case the Reynolds analogy hold exactly and the
heat transfer coefficient can be computed as
\begin{equation}
h = \frac{\tau_w C_p}{\bar{u}}\,.
\label{reynolds-analogy}
\end{equation}

For air the Prandtl number, $Pr_{\mathrm{air}} = 0.71$, so the Reynolds
analogy is considered a reasonable approximation \citep[p.~410]{kreith}.  A
number of empirical relations have been developed to provide more accurate
approximations for specific geometries and Prandtl and Reynolds number
ranges. These have not been considered in our paper.

Using the wall shear stress values computed for the Piccadilly Line tunnels in
this section and appropriate constants for air, the heat transfer coefficients
for open tunnel and annular flow can be computed
using~\eqref{reynolds-analogy} as
\begin{itemize}
\item Open tunnel flow: $h=44$~W/m$^2\cdot$K
\item Annulus flow: $h=110$~W/m$^2\cdot$K
\end{itemize}
Details of the parameters assumed and computations are given in
Appendix~\ref{apxA}.  Notice the large increase in the coefficient for the
annular flow case, due to the high shear generated by passing
trains.

\section{Soil Temperature Response to an Impulse Change in Air Temperature}
\label{sec:conduction1}

In this and the next section we consider the conduction of heat through the
soil surrounding a tunnel of circular cross section with radius $a$. The
temperature of the soil is governed by the one dimensional heat equation in
radial coodinates,
\begin{equation}
\frac{\partial{T(r,t)}}{\partial{t}} = \kappa \left(\frac{\partial^2{T(r,t)}}{\partial{r^2}}+\frac{1}{r} \frac{\partial{T}}{\partial{r}}\right)\,.
\label{heat equation}
\end{equation}
Here $r$ is the radial coordinate, $t$ is time, $T(r,t)$ is the temperature of
the solid in $a \leq r \leq \infty$ and $\kappa = K/ \rho C$ is the thermal
diffusivity of the solid, where $K$, $\rho$ and $C$ are the conductivity,
density and specific heat capacity respectively.

\begin{figure}[h!]
\begin{center}
\includegraphics[scale=0.8]{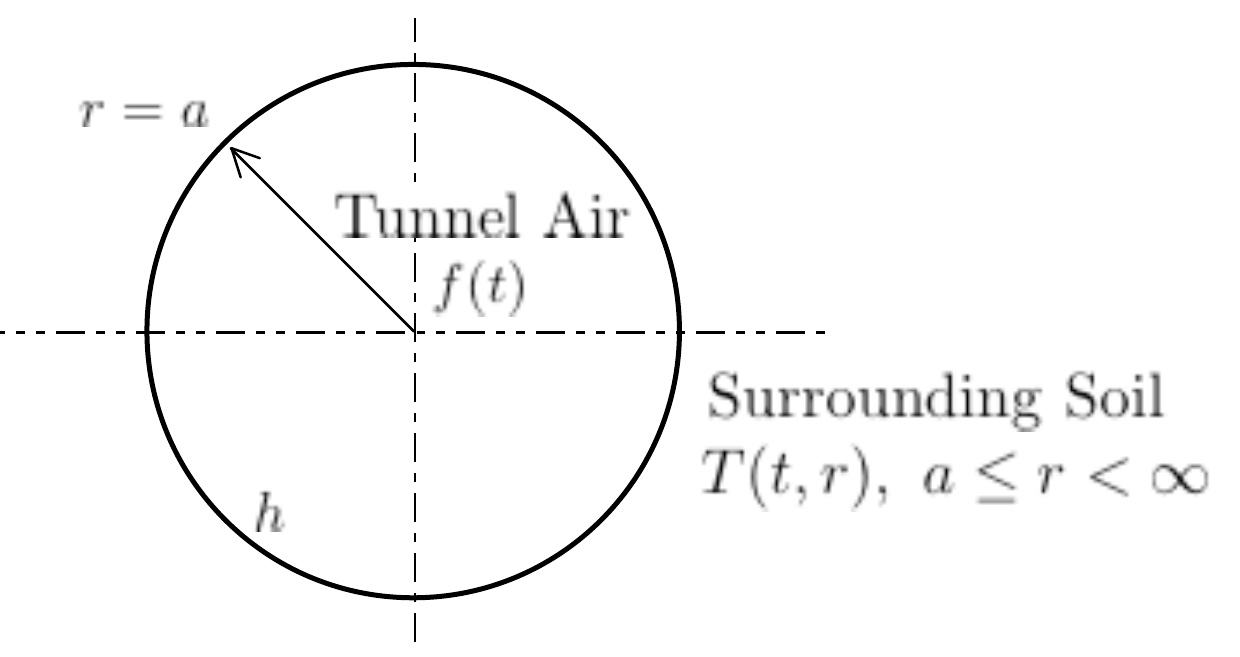}
\caption{Tunnel of radius $r=a$, air temperature $f(t)$, heat transfer
  coefficient $h$.}
\label{figure_layout}
\end{center}
\end{figure}

A number of assumptions have been made in the above formulation of the one
dimensional problem.  Firstly, a single cross section of tunnel has been
considered meaning that any heat conduction along the axis of the tunnel
(perpendicular to the radial coordinate) is neglected. Secondly, it has been
assumed that there is no angular dependence to the temperature profile; that
there is not top or bottom to the tunnel.  This infers that the tunnel wall is
an infinite distance from any other heat source or boundary.  The validity of
these assumptions is discussed in more detail later later in this section. The
geometry of the problem under consideration is shown in
Figure~\ref{figure_layout}.

Conductive heat transfer through the solid is driven by the difference between
the air temperature within the tunnel, $f(t)$, and the surface temperature of
the tunnel, $T(a,t)$.  Convective heat transfer between the air and wall obeys
Newton's cooling law. Heat flux at the tunnel wall is continuous, which
requires
\begin{equation}
- \frac{\partial{T(a,t)}}{\partial r} = H \left(f(t) - T(a,t)\right)~,
\label{boundary condition}
\end{equation}
where $H = h/K$, which is taken as a constant in this section (in Section~\ref{sec:fluctuations}  we
investigate the effects of the short-term oscillations in $h$ due to passing trains).  The boundary condition at infinity is
\begin{equation}
T(\infty,t)~ \text{finite as}~r \rightarrow \infty\,.
\label{infinity condition}
\end{equation}

For ease of computation we have taken $f(t)$ as the absolute temperature
difference between the tunnel air and the `deep sink' temperature of the
soil. Hence $T(\infty,t) = 0$, and $T(r,t)$ is the absolute temperature
difference between the soil temperature within the thermal boundary layer
around the tunnel and the constant `deep sink' soil temperature.

Owing to the linearity of the problem, the temperature of the air in the
tunnel, $f(t)$, can be expressed as the sum of any number of components. We
have assumed that a sum of a generic steady and periodic component $f(t) = f_s
+ f_p(t)$. Physically the steady component is due to heat sources within the
tunnel environment elevating the air temperature. The periodic components
result from daily and seasonal fluctuations in ambient air temperature which
make their way into the tunnel.

In this section we are investigate the behavior of temperature in the soil
due to an instantaneous change in air temperature.  The initial temperature of
the soil is assumed constant, $T(r,0)=0$ for $a \leq r < \infty$---the tunnel
wall and all of the surrounding soil are initially at the `deep sink'
temperature. The constant component of the air temperature is assumed to be
$f_s=1$. Both of these assumptions can be made without loss of generality.

The long-term solution to this problem is the steady state $T(r) = f_s$ as $t
\rightarrow \infty$. Here all the spatial and temporal derivatives eventually
go to zero and hence the soil temperature approaches the air temperature. This
steady state solution is sufficient where only the long-term behavior of the
tunnel wall temperature is required. However, a solution of the transient
behaviour is useful in that it gives the time needed to reach the steady
state. This is of interest, for instance, in determining how long an
underground rail system will operate before initial transients die out and
peak temperatures are reached .

\subsection{Solution of the Equation}

Application of the Laplace transform to the governing equation \eqref{heat
equation} results in the subsidiary ordinary differential equation
\begin{equation}
\frac{d^2 \bar{T}}{d r^2} + \frac{1}{r} \frac{d \bar{T}}{dr} - q^2 \bar{T} =
0\,,
\label{subsidiary equation}
\end{equation}
where $p$ is the transform space, $\bar{T}$ is the transformation of $T$ and
$q^2 = p / \kappa$. This takes the form of the modified Bessel equation of
order zero which has solution \citep{HMF}
\begin{equation}
\bar{T} = B K_0(q r)\,,
\label{modified bessel solution general}
\end{equation}
where $B$ is a constants and the other linearly independent solution is ruled
out by the condition at infinity~\eqref{infinity condition}.  $B$ is
determined from the Laplace transform of the boundary
condition~\eqref{boundary condition},
\begin{equation}
\bar{T}(r,t) = \frac{H K_0(q r)}{p\left(H K_0(q a) - q K_1(q a)\right)}\,.
\label{laplace solution}
\end{equation}

Applying the inverse Laplace transform to~\eqref{laplace solution},
\begin{equation}
T(r,t) = \frac{H}{2 \pi i} \int_{\gamma - i \infty}^{\gamma + i \infty} \exp(\lambda t) \frac{K_0(\mu r)}{\lambda \left(H K_0(\mu a) - \mu K_1(\mu a)\right)} d \lambda\,,
\end{equation}
where $\mu = \sqrt{\lambda / \kappa}$. The integral above contains a branch
point at $\lambda = 0$. The contour integral is taken over the usual
``keyhole'' contour.

We separate the real and imaginary components of the
integral using the relationship
\begin{equation}
K_0\left(x \exp( i \pi / 2)\right) = -\frac{1}{2} i \pi \left(J_0(x) - i\, Y_0(x)\right)\, ,
\end{equation}
and find
\begin{multline}
T(r,t) =
\frac{2 H}{\pi}
\int_0^\infty \frac{du}{u} \,\exp(-\kappa u^2 t) \\ \times
\frac{J_0(u r)\left(u Y_1(ua)+HY_0(ua)\right) - i\, Y_0(ur)\left(u J_1(ua)+HJ_0(ua)\right)}
{\left(u J_1(ua)+HJ_0(ua)\right)^2 + \left(u Y_1(ua)+HY_0(ua)\right)^2} 
\,.
\label{transient solution}
\end{multline}
The solution presented above is adapted from the outline given in \citet[\S 13.5]{carslaw}.

Substituting appropriate parameters into \eqref{transient solution} gives
results for the Piccadilly Line tunnels. Here
the tunnel radius, $a=1.7$~m, thermal conductivity of the soil is
$K=0.35$~W/(m$\cdot$K) and the heat transfer coefficient $h=44$ and
$h=110$~W/m$^2\cdot$K for open tunnel and annulus flow
respectively. 

\begin{figure}[h!]
\begin{center}
\includegraphics[scale=0.5]{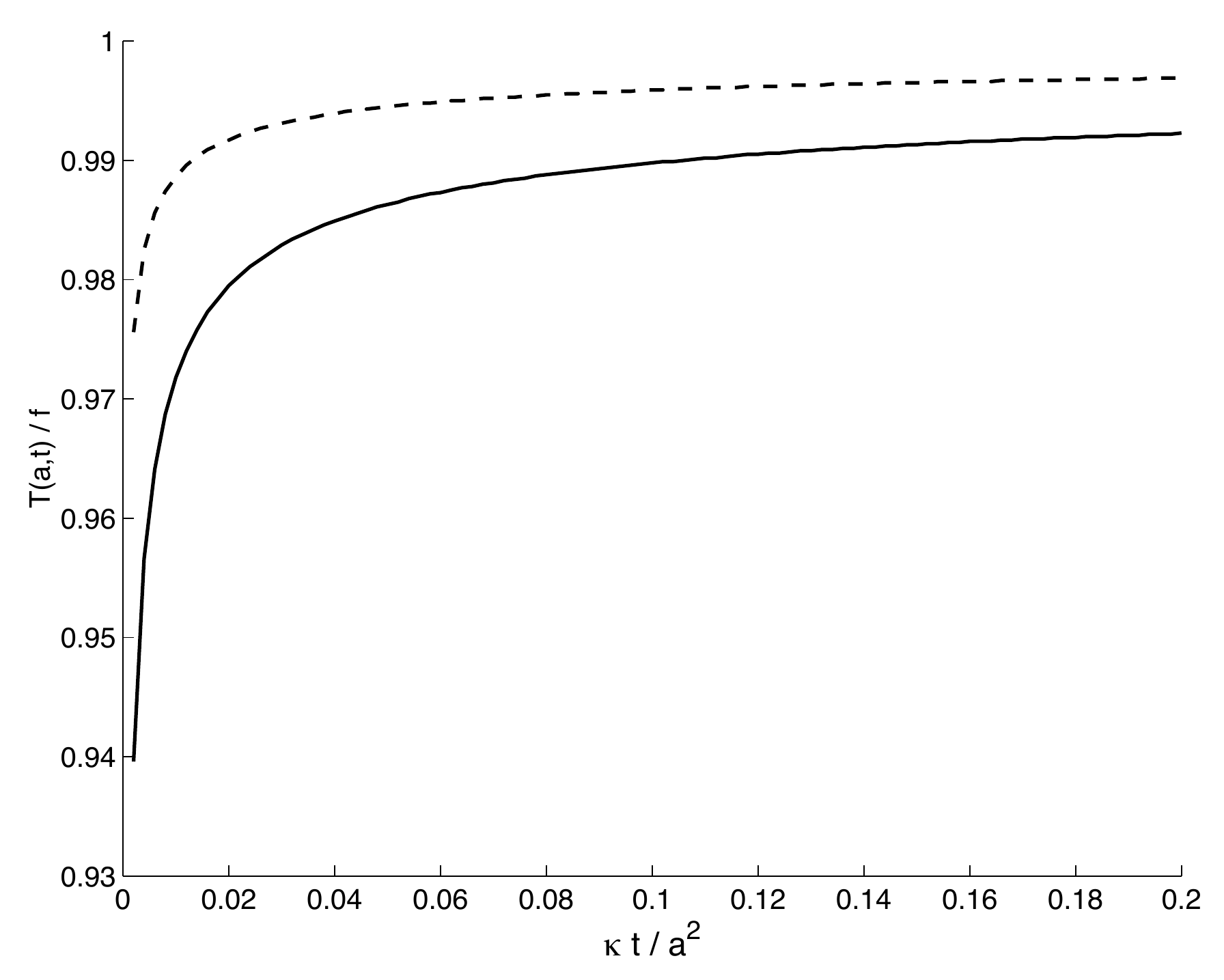}
\caption{Tunnel wall surface temperature profiles for  $h = 44(-), 110(-~-)$.}
\label{graph_picc_transient}
\end{center}
\end{figure}
Figure~\ref{graph_picc_transient} shows the behavior of the wall temperature
against time for the two heat transfer coefficients for an impulse change in
air temperature. Temperature at the wall surface approaches the air
temperature faster for a larger value of $H$.  The profiles for $h=44$ and
$110$ reach 0.99 of the asymptotic value when $\kappa t / a^2 =$ 0.2 and 0.02
respectively. This in turn corresponds to times of approximately 5 and 50 days
respectively (see Appendix \ref{apxA} for calculations).

\section{Soil Temperature Response to Periodic Variations in Air Temperature}
\label{sec:conduction2}

As discussed in Section \ref{sec:conduction1}, we write difference between the
ground deep sink temperature and the air at any location in an underground
tunnel as a steady and time-dependent component, $f(t)=f_s+f_p(t)$.  The time
varying component, due to diurnal and seasonal fluctuations in ambient
temperature, is expected to be a periodic function of time.  The response of
the radial soil temperature profile $T(t,r)$ to the steady component was
discussed in the previous section.  In this section we deal with the
fluctuating component.

A solution to the heat equation with $f_p(t)$ a periodic function of time can
be found using the Laplace transform techniques used in the previous
section. As was the case there, a decaying transient term will result from a
branch point. There will also be two simple poles within the contour which can
be evaluated using Cauchy's Theorem.  These simple poles give the limit cycle
of the steady oscillating solution to the problem.  The steady oscillating
component of the solution is of greatest interest, as it describes the
long-term behavior of temperature within the soil surrounding the tunnel, and
hence heat flux through the walls.

In this section the limit cycle is found using assumptions about the form of
the cycle rather than a Laplace transform. The Laplace transform is necessary
to find the transient component of the solution associated with the
oscillating solution, which is not considered here.

\subsection{General Form of the Oscillating Solution}

It is expected that the periodic components of the tunnel air temperature,
$f_p(t)$, will produce periodic fluctuations of the same frequency in the
temperature of the solid. Hence, the temperature field in the solid is
expected to take the form
\begin{equation}
T(r,t) = e^{i \omega t}R(r)\,,
\label{separation}
\end{equation}
where $\omega$ is the frequency of fluctuation in $f_p(t)$ and $R(r)$ is independent of time. Substituting this form into~\eqref{heat equation} gives
\begin{equation}
r^2 \frac{d^2R}{dr^2} + r \frac{dR}{dr} -\frac{i \omega}{\kappa} r^2 R = 0\,,
\end{equation}
the modified Bessel's equation of order zero. This has solution given by
\eqref{modified bessel solution general} with $q=\sqrt{{i
\omega}/{\kappa}}$ and $B$ is a constant, and once again the boundary
condition~\eqref{infinity condition} has been applied. Substituting the solution~\eqref{modified bessel
  solution general} into the boundary condition at the
tunnel wall~\eqref{boundary condition} yields the constant $B$, hence
\begin{equation}
T(r,t) = \frac{H f_p(t) K_0\left(\sqrt{\frac{i \omega}{\kappa}}r\right)}
{H K_0\left(\sqrt{\frac{i \omega}{\kappa}}a\right) + 
\sqrt{\frac{i \omega}{\kappa}} K_1\left(\sqrt{\frac{i \omega}{\kappa}}a\right)}\, .
\label{bessel form}
\end{equation}

In order to extract the real component of the solution, the imaginary terms in
the denominator and in the arguments of the modified Bessel functions must be
separated.  This is achieved by writing the modified Bessel function in
terms of the real and imaginary components of the Kelvin function,
$\mathrm{ker}_\nu$ and $\mathrm{kei}_\nu$ using the relation
\begin{equation}
e^{-\nu \pi i / 2} K_\nu (x e^{\pi i / 4}) = \mathrm{ker}_\nu (x) +  i\,  \mathrm{kei}_\nu (x)\,.
\label{bessel kelvin}
\end{equation}

The final relation can be greatly simplified when expressed in terms of the Kelvin function modulus, $N_\nu$, and phase, $\phi_\nu$, defined as
\begin{gather}
N_\nu(x) = \sqrt{\mathrm{ker}_\nu^2(x) + \mathrm{kei}_\nu^2(x)}\,,
\label{modulus}\\
\phi_\nu(x) = \arctan \left(\mathrm{kei}_\nu(x) / \mathrm{ker}_\nu(x)\right)\,.
\label{phase}
\end{gather}
The relations \eqref{bessel kelvin}, \eqref{modulus} and \eqref{phase} are
given in \citet{HMF}.

Rewriting \eqref{bessel form} in terms of the modulus and phase of Kelvin
functions gives
\begin{equation}
T(r,t) = \frac{H f_p(t) N_0(\omega'r) \exp\left(i \phi_0(\omega'r)\right)}
{H N_0(\omega'a) \exp\left(i \phi_0(\omega'a)\right)+\sqrt{\frac{i\, \omega}{\kappa}}a N_1(\omega'a) \exp \left(i\left(\phi_1(\omega'a)+ \frac{\pi}{2}\right)\right)}\,,
\label{mod phase form}
\end{equation}
where $\omega' =  \sqrt{\omega / \kappa}$ is the inverse length scale to which
the oscillating temperature variations penetrate the solid.

Specifying the form of the periodic air temperature variations to be $f_p(t) = \sin(\omega t + \epsilon)$, which is still of a general enough form to accommodate the desired fluctuations, helps to simplify the expression further.

Using Euler's formula and the harmonic addition theorem, the denominator of \eqref{mod phase form} can be separated into real and imaginary parts ($\alpha + i\,\beta$). Multiplying numerator and denominator by the complex conjugate of this gives
\begin{equation}
T(r,t) = \frac{H N_0(\omega'r) 
\sin \left(\omega t + \epsilon + \phi_0(\omega'r)-\arctan(\beta/\alpha)\right)}
{\sqrt{\alpha^2 + \beta^2}}\,,
\label{periodic solution}
\end{equation}
where
\begin{multline}
\alpha = H N_0(\omega'a) \cos \left(\phi_0(\omega'a)\right) \\ +
\frac{\omega'}
{\sqrt{2}} N_1(\omega'a)\left(\cos\left(\phi_1(\omega'a)+\pi /2\right)-\sin\left(\phi_1(\omega'a)+\pi /2\right)\right),
\label{alpha}
\end{multline}
\begin{multline}
\beta = H N_0(\omega'a) \sin(\phi_0(\omega'a)) \\ +
\frac{\omega'}
{\sqrt{2}} N_1(\omega'a)\left(\cos \left(\phi_1(\omega'a)+\pi /2\right)+\sin \left(\phi_1(\omega'a)+\pi /2\right)\right).
\label{beta}
\end{multline}

The fact that the analytic solution to the problem \eqref{periodic solution}
is stated in terms of the modulus and phase of Kelvin functions means that its
behaviour may not be immediately apparent.  The graphs below illustrate its
general features.  In all the computations we used unity for all parameters
other than $\omega$, which was set to $2 \pi$, so that $f_p(t)$ has period one
and $\epsilon$, which was set to zero. The Kelvin functions have been computed
using asymptotic approximations and code given in \citet{CSF}.

\begin{figure}[h!]
\begin{center}
\includegraphics[scale=0.5]{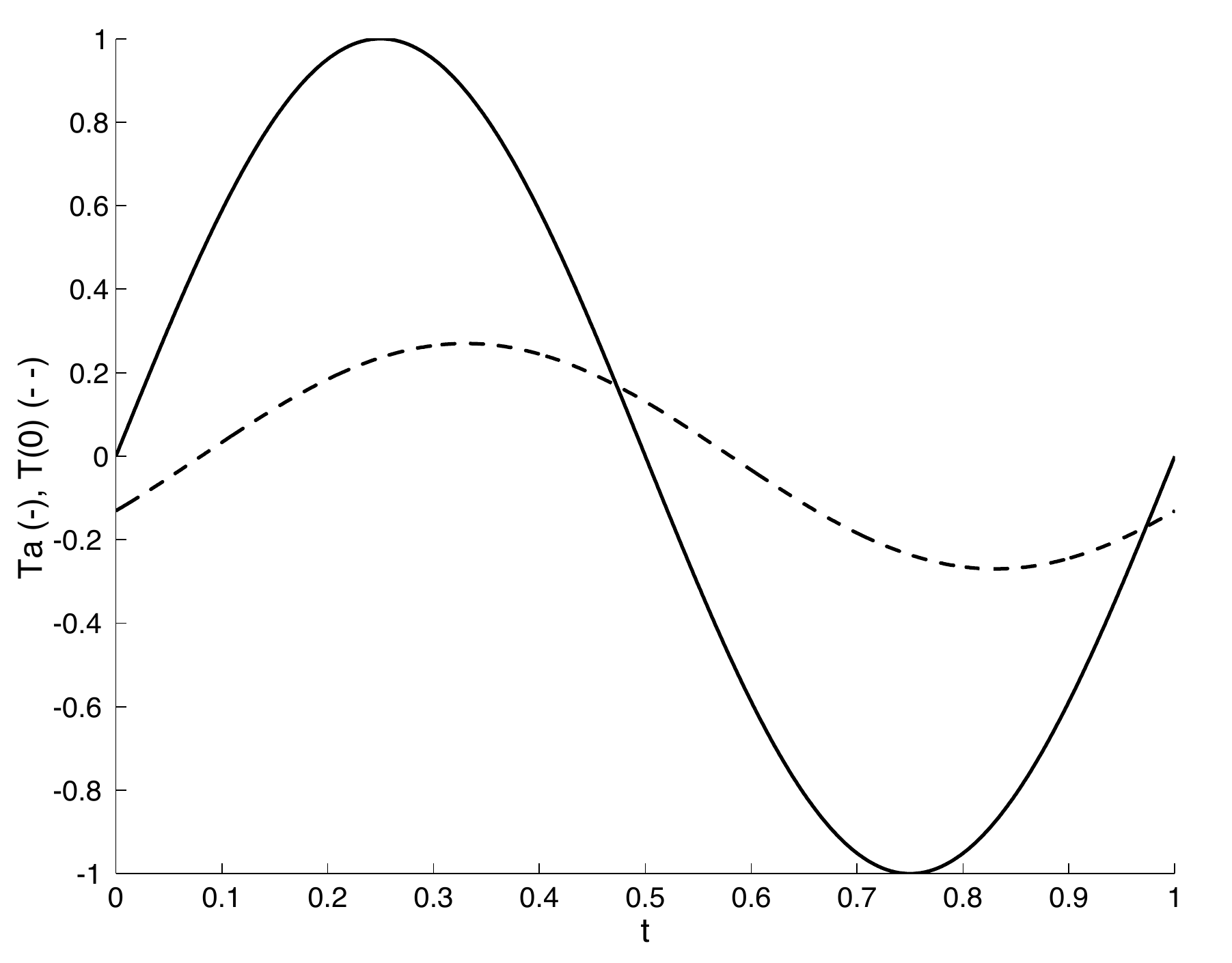}
\caption{Air (--) and tunnel wall (-~-) temperatures vs time.}
\label{graph_temps}
\end{center}
\end{figure}

Figure~\ref{graph_temps} shows the air, $f_p(t) = \sin(2 \pi t)$, and tunnel
surface temperature, $T(a,t)$, plotted against time for one period of
oscillation. The temperature at the wall surface has the same period as the
air temperature, but has a smaller amplitude and is out of phase with
(slightly lagging) the wall temperature.

\begin{figure}[h!]
\begin{center}
\includegraphics[scale=0.5]{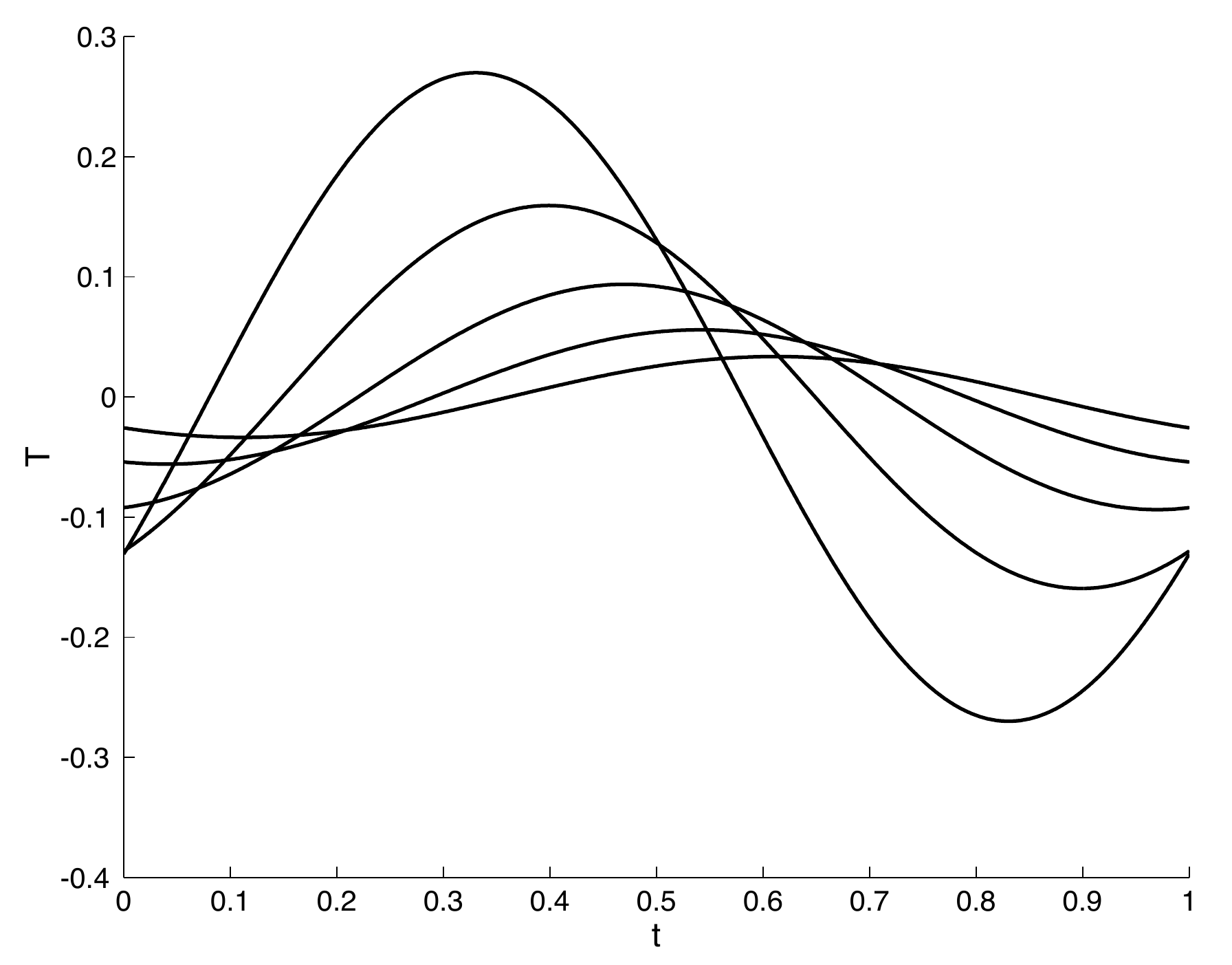}
\caption{Temporal variation of temperature various depths: $r = a$, and $a +
0.25, 0.5, 0.75, 1, 1.5, 2$.}
\label{graph_tempdepth}
\end{center}
\end{figure}

Figure~\ref{graph_tempdepth} is a plot of temperature against time at the tunnel surface and at progressive depths into the solid. This plot also shows that there is a decay in magnitude of the temperature oscillations as we move into the solid and a continual shift in the phase of the temperature oscillations.

The general form of the solution \eqref{periodic solution} has three distinct
components relating to the amplitude of the tunnel wall temperature, the decay
of the temperature with radial distance from the tunnel wall and the phase of
temperature oscillations through a soil thermal boundary layer when compared
to the air temperature oscillations.

The amplitude component of the solution, including the radial component of the
solution evaluated at the tunnel wall is
\begin{equation}
\frac{H N_0(\omega' a)}{\sqrt{\alpha^2+\beta^2}}\,,
\label{eq:amplitude}
\end{equation}
where $\alpha$ and $\beta$ are given by~\eqref{alpha} and~\eqref{beta}
respectively. Although this term is a constant for any given problem, it will
vary with the value of the parameters: $a, H=h/K$ and $ \omega' = \sqrt{\omega
/ \kappa}$.

The amplitude component~\eqref{eq:amplitude} plotted against the parameter
$h/K$ is shown in Figure~\ref{graph_const_term} for typical values for the
Piccadilly line tunnels (see Appendix \ref{apxA}).  The two profiles plotted
are for oscillations with periods of one day and one year.  As $h/K$
increases, the constant term asymptotically approaches unity. That is, the
temperature at the wall surface $r=a$ approaches the oscillating air
temperature. If the convective heat transfer coefficient, $h$, is very large
or if the thermal conductivity of the solid, $K$, is very small, then the
oscillations in the air temperature are also seen in the wall temperature.
Further, this approach for increasing $h/K$ is dependent on the frequency of
oscillation, $\omega$. For the same value of $h/K$, the wall surface
temperature approaches the air temperature more quickly for lower frequency
oscillations. This is shown in Figure~(\ref{graph_const_term}), as the
approach for yearly oscillations is more pronounced than for daily
oscillations.
\begin{figure}[h!]
\begin{center}
\includegraphics[scale=0.5]{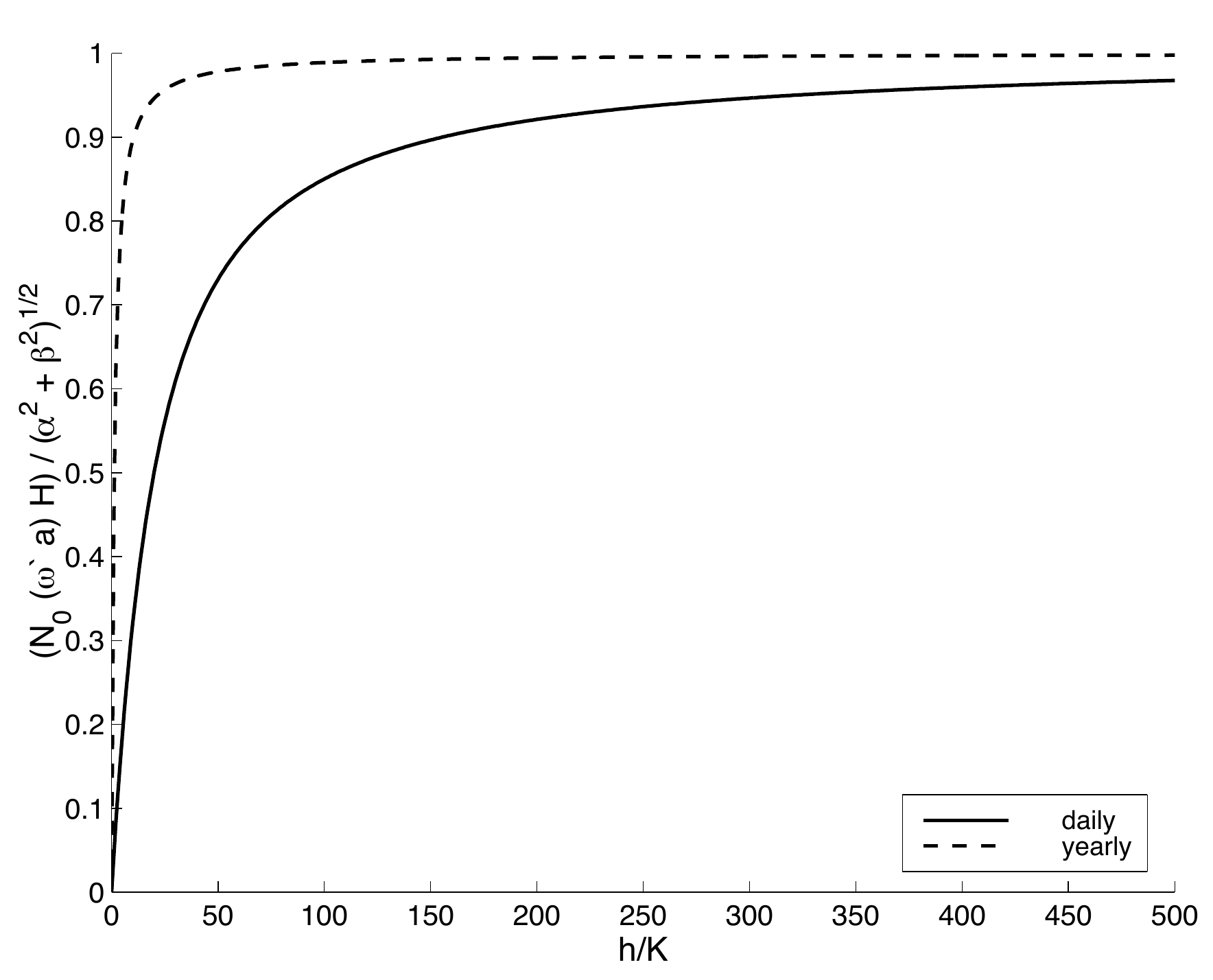}
\caption{Amplitude component~\eqref{eq:amplitude}.}
\label{graph_const_term}
\end{center}
\end{figure}

\begin{figure}[h!]
\begin{center}
\includegraphics[scale=0.5]{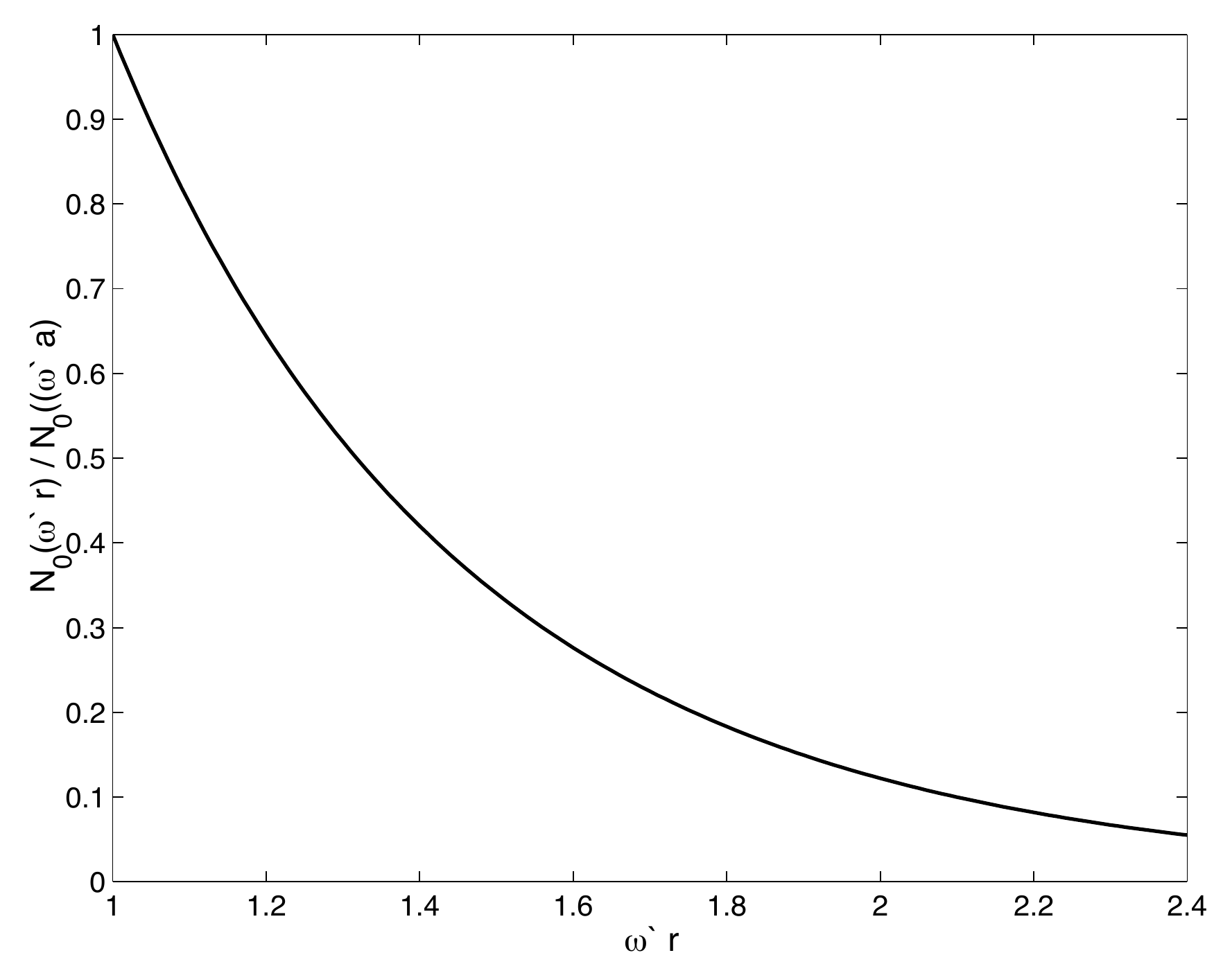}
\caption{Kelvin modulus~$N_0(\sqrt{\omega / \kappa}r)$, showing the decay of
temperature with distance from wall.}
\label{graph_N}
\end{center}
\end{figure}

The radial component of the periodic solution is $N_0(\omega' r)$. This
reduces the magnitude of any temperature variation with distance from the
tunnel wall in proportion to $\sqrt{\omega'}$ defining the depth of a thermal
boundary layer around the tunnel which is affected by oscillating air
temperatures within the tunnel. This is shown in Figure~\ref{graph_N}, a plot
of $N_0(\sqrt{\omega / \kappa}r)$ against time normalised to $N_0(\sqrt{\omega
/ \kappa}a)$.
The magnitude of the temperature fluctuations has reduced to approximately 0.1
that of the fluctuations at the wall surface when $\omega` r = 2.2$. Using
parameters for the Piccadilly line tunnels, diurnal temperature fluctuations
die out to 0.1 of the peak value at the tunnel wall in approximately 0.1
meters. Yearly fluctuations die out within approximately 1.8 meters of the
tunnel wall (see Appendix A for calculations).  For the analogous problem in
Cartesian coordinates, this decay with distance from a boundary is
exponential, $\exp(-\omega x / \sqrt{2})$ \citep[\S 2.6]{carslaw}.

The periodic component of the solution for an air temperature of
$\sin(\omega t + \epsilon)$ is
\begin{equation}
 \sin \left(\  t + \epsilon + \phi_0(\omega'r)-\arctan(\beta/\alpha)\right)\,.
\end{equation}

The two additional arguments of the periodic function change the phase of the
temperature oscillations in the solid. In general, the oscillations at the
tunnel wall will be out of phase with the oscillations in air temperature. The
phase difference will also change with distance from the wall by the component
$\phi_0(\omega'r)$.  This effect is shown in
Figure~(\ref{graph_tempdepth}). As we move further into the solid the
temperature oscillations not only reduce in amplitude, but also experience a
continual phase lag, also a function of the oscillating frequency. This effect
has been measured in the soil surrounding London Underground tunnels---a plot
of yearly oscillations in soil temperature at increasing distance from a
tunnel wall is given in \citet[Fig. 2]{Ampofo}.

\subsection{Heat Flux through the Tunnel Wall}

We have presented the main results of this section in terms of the oscillating
temperature fields that form a thermal boundary layer around the
tunnel. Computationally, however, these results are more useful when expressed
in terms of the heat flux through the tunnel wall, $\dot{q}(r,t)$, defined as
\begin{equation}
\dot{q}(r,t) = -K\frac{\partial T(r,t)}{\partial r}\, 
\end{equation}
where we take the spatial derivative of the Bessel form of the solution,
\eqref{bessel form}, containing both real and imaginary parts as both will
contribute to the real part of the flux term. This yields
\begin{multline}
\dot{q}(r,t) = \frac{h f_p(t) N_1(\omega' a)}{\sqrt{\alpha^2+\beta^2}} 
\biggl[\cos \left(\phi_1(\omega' a) + \pi / 2 + \arctan
    \left(\frac{\alpha-\beta}{\alpha + \beta}\right)\right) \\ + 
i\, \left(\cos(\phi_1(\omega' a) + \pi / 2 - \arctan \left(\frac{\alpha+\beta}{\alpha - \beta}\right)\right)\biggr]  \,.
\label{periodic flux preliminary}
\end{multline}
We write the right-hand side of~\eqref{periodic flux preliminary} as
$\dot{q}_{\mathrm{r}}+i\,\dot{q}_{\mathrm{i}}$. Both the real and imaginary
parts are used in calculations in Section \ref{sec:fluctuations}, where the
coupled relationship between tunnel air and surface temperatures is
investigated. The real component $\dot{q}_{\mathrm{r}}$ is only required for
the explicit form of the flux as given below.  For the case where $f_p(t)=
\sin(\omega t + \epsilon)$, the final relationship becomes
\begin{equation}
\dot{q}(a,t)
= \frac{h N_1(\omega' a)}{\sqrt{\alpha^2+\beta^2}}\, \cos \left(\omega t +\epsilon + \phi_1(\omega' a) + \arctan \left(\frac{\alpha-\beta}{\alpha + \beta}\right)\right)
\,.
\label{periodic flux}
\end{equation}

With this we have arrived at an explicit relationship for the heat flux into
the soil surrounding a cross section of tunnel due to periodic oscillations in
the tunnel air temperature. In practice this quantifies the cooling effect
that the thermal mass of the soil can provide by rejecting heat during the
night or the winter and absorbing it during the day or the summer. In the next
Section we investigate the effect of the rapidly varying heat transfer
coefficient due to passing trains on this result, which assumed $h$ is a
constant.

\section{Soil Temperature Response to Rapidly Fluctuating Heat Transfer Coefficient}
\label{sec:fluctuations}
In this section we consider the effects of varying heat transfer coefficient
$h$ as described in Section~\ref{sec:turbulentflow2}. These fluctuations are
caused by train movements and are treated as events that occur over a time
period of several minutes, the effects of which have been ignored in the
previous two sections.

Throughout this and subsequent sections an averaging function, denoted by
$\langle \cdot \rangle$, over the short time period $\varepsilon$ is defined
as
\begin{equation}
\langle \cdot \rangle = \frac{1}{\varepsilon} \int_0^\varepsilon \cdot~dt\,.
\label{average operator}
\end{equation}

This section is divided into three parts.  In Section~\ref{sec:indep} we
examine the standard assumptions that temperature in a tunnel does not
fluctuate over short time scales.  In Section~\ref{sec:notindep} we drop this
assumption and show that the long-time heat flux is modified by short-term
correlations between the heat transfer coefficient and the change of
temperature in the tunnel. 

\subsection{Fluctuations Independent of Air Temperature}
\label{sec:indep}

A first approximation of the effect of a rapidly fluctuating heat transfer
coefficient can be made by assuming that temperature fluctuations occur
independently of any changes in tunnel air temperature and on a much shorter
time scale. It is typically assumed that all fluctuations in tunnel air
temperature have a period of either one day or one year and are due to diurnal
and seasonal fluctuations in external air temperature drawn into the tunnels
by the train piston effect.  Assuming a typical train headway (time between
subsequent trains) of two minutes ($\varepsilon = 120$ seconds), the periods
of daily and yearly oscillations expected in the tunnel air ($f_\varepsilon
(t)$) and surrounding soil ($T(r,t)$) are larger by factors of approximately
$10^4$ and $10^7$.

As any changes in air and soil temperature occur over a period much longer
than $\varepsilon$, the averaging operator~(\ref{average operator}) has no
effect, that is $\langle f_\varepsilon (t) \rangle = f_\varepsilon (t)$ and
$\langle T(r,t) \rangle = T(r,t)$, and the governing heat equation~(\ref{heat
equation}) is invariant under averaging. Averaging the boundary
condition~\eqref{boundary condition} gives the relationship
\begin{equation}
- \frac{\partial T(a,t)}{\partial r} = \langle H(t) \rangle \bigl(f_\varepsilon (t) - T(a,t) \bigr).
\end{equation}
Under these assumptions the heat transfer coefficient can simply
be time averaged over a whole number of oscillations and treated as a
constant. It follows that the results presented in the previous sections hold
for a time averaged heat transfer coefficient $\langle h(t) \rangle$.  This
approach is generally used engineering calculations such as \citet{barrow}.

\subsection{Including Short-Term Fluctuations in Air Temperature}
\label{sec:notindep}

The assumptions detailed above are, however, flawed. As a train approaches a
station and brakes, the kinetic energy possessed by the train and passengers
is converted to heat energy. Train-generated heat accounts for as much as
eighty five percent of the heat load within a typical London Underground
station \citep{UC}. It is therefore reasonable to assume that temperatures
within underground rail tunnels will fluctuate, to some extent, over the same
short time scales at which the heat transfer varies as both are caused by the
passing of a train.

A better assumption is therefore that the tunnel air temperature does vary
over the time scale $\varepsilon$, so $\langle f_\varepsilon (t) \rangle \neq
f_\varepsilon (t)$. However, the large thermal inertia associated with the
tunnel walls means that the wall surface temperature $T(a,t)$ will not be
affected on this short time scale. This can be shown using the previous
analytic solution \eqref{periodic solution} for an oscillating air
temperature with $\omega = \varepsilon = 2\pi/120$ radians per second.

Under the above assumptions the boundary condition \eqref{boundary condition}
becomes
\begin{equation}
- \frac{\partial T(a,t)}{\partial r} = \langle H(t) f_\varepsilon(t) \rangle -
  \langle H(t) \rangle T(a,t)\,.
\end{equation}
Hence, the solution to the heat equation given as (\ref{periodic flux
preliminary}) becomes
\begin{equation}
\langle \dot{q}(a,t) \rangle
= \frac{K \langle H(t) f_\varepsilon (t) \rangle N_1(\omega'
  a)}{\sqrt{\alpha^2+\beta^2}}\,
\cos \left(\phi_1(\omega' a) + \arctan \left(\frac{\alpha-\beta}{\alpha +
  \beta}\right)\right)
\label{averaged flux}
\end{equation}
for the short-term average where $\alpha$ and $\beta$ now contain $\langle H
\rangle$ rather than $H$. Note that if the short-term fluctuations in
$f_\varepsilon (t)$ are ignored the heat flux~\eqref{averaged flux} vanishes.
If $H(t)$ and $f_\varepsilon (t)$ are statistically independent of each other,
then $\langle H(t) f_\varepsilon (t) \rangle = \langle H(t) \rangle \langle
f_\varepsilon (t) \rangle$, and the short-term average of each function is all
that is required. However, this is not the case as the short-term fluctuations
in both $H(t)$ and $f_\varepsilon (t)$ are both caused by the passage of
trains through a tunnel.

As discussed in Section \ref{sec:conduction2}, the heat transfer coefficient
$H(t)$ can be assumed to take a base value of $H_1$ when there is a train
passing a particular point in a tunnel, and a lower value of $H_2$ when a
train is not present. Although the air flows before and after the passage of
the train are likely to be complicated, $H(t)$ can be approximated by a step
function.  Normalising the short-term fluctuations ($\varepsilon = 1$), and
defining the fraction of the train service headway for which a train is
passing any given point in a tunnel as $\sigma$, which can be computed from
the train headway, average speed and length (see Appendix \ref{apxA} for
assumed values and computations), then
\begin{equation}
H(t) = \begin{cases} H_1\,, \quad & \text{for } 0 < t < \sigma\,; \\ 
H_2\,,&\text{for } \sigma < t < 1\,.
\end{cases}
\end{equation}

The short-term behavior of the air temperature in a tunnel as a train passes
is somewhat more difficult to predict. Detailed output from a numerical
simulation can be used or measurements made in an existing tunnel to confirm
any assumptions or simulations.  For ease of computation, however, we have not
unrealistically assumed that these fluctuations in air temperature are
sinusoidal with a generic phase lead or lag, $\varphi$, to show the effects of
$H(t)$ and $f_\varepsilon (t)$ being in or out of phase with each other:
\begin{equation}
f_\varepsilon (t) = \sin(2 \pi t / \varepsilon + \varphi).
\end{equation}
A more complicated form of $f_\varepsilon(t)$ can be represented by a Fourier
series. All subsequent calculations would then be essentially the same as
those presented here.  The averaged forcing term is
\begin{equation}
\langle H(t) f_\varepsilon (t) \rangle = H_1 \int_0^\sigma \sin(2 \pi t + \varphi) dt + H_2 \int_\sigma^1 \sin(2 \pi t + \varphi) dt \,.
\end{equation}
Hence
\begin{equation}
\langle H(t) f_\varepsilon(t) \rangle = \frac{1}{\pi} (H_1-H_2) \sin(\pi \sigma) \sin(\pi \sigma + \varphi)\,.
\end{equation}

Calculations have been carried out using the values for $H_1$, $H_2$,
$\sigma$, and $\varepsilon$ from the Piccadilly line parameters given in
Appendix \ref{apxA}.  A short-term temperature oscillation with a magnitude of
one degree centigrade has been assumed.  This is an estimate, and detailed
measurements or simulations are required for any particular application.

The averaged forcing term is a sinusoidal function of the phase difference
between $H(t)$ and $f_\varepsilon (t)$. For the parameters above the amplitude
is approximately 18.5 for a relatively small fluctuation in air temperature.
Physically, this means that heat transfer into the wall is increased if the
air temperature is warmer when a train is present and the heat transfer
coefficient is greater. As the train is a heat source, we expect this to be
the case. Hence, ignoring the correlation between short-term oscillations in
air temperature and heat transfer coefficient will underestimate heat transfer
to the wall.

\section{Heat Transfer from a Tunnel of Finite Length}
\label{sec:tunnel}

Analytic solutions presented thus far have concentrated on a single location
in a tunnel. In all circumstances a generic air temperature, $f(t)$, has been
considered. However, in an underground environment the temperature of the air
and surrounding soil are coupled. If there is heat flux from the air into the
tunnel wall the air temperature is cooled.

\begin{figure}[h!]
\begin{center}
\includegraphics[scale=0.8]{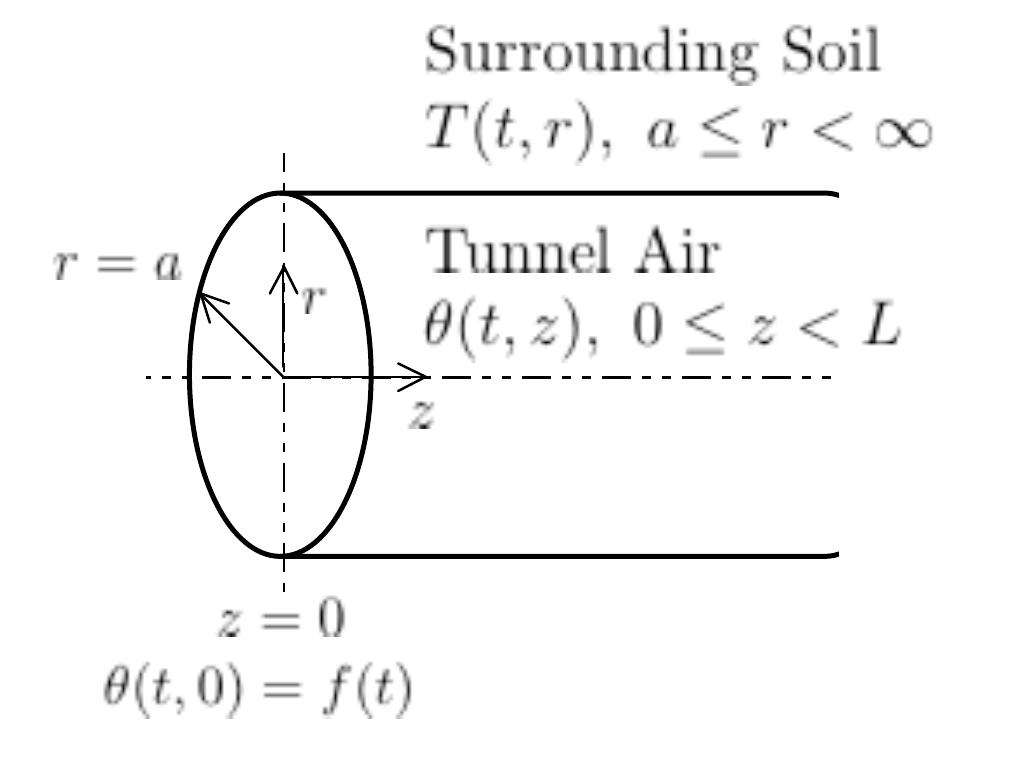}
\caption{Tunnel of radius $r=a$ and length $z=L$, air temperature
  $\theta(t,z)$.}
\label{figure_tunnel}
\end{center}
\end{figure}

In this section air temperature within a tunnel, $\theta(t,z)$, is assumed to
be a function of both time and the axial distance along the tunnel $z$. It is
assumed that the temperature at the beginning of the tunnel is known and is a
simple harmonic function of time,
\begin{equation}
  \theta(t,0) = f(t) = D \exp(i \omega t).
  \label{eq:BCz0}
\end{equation}
A version of the solution presented here is given in \citet{peavy}, where the
homogeneous problem is considered. This does not allow for the heat flux due
to short-term fluctuations, or any other inhomogeneous heat generation within
the tunnel, so is of limited use.

The equation governing the temperature of air flowing axially through a
tunnel, $\theta(t,z)$, where $z$ is the axial coordinate is
\begin{equation}
M C_p \left(\frac{\partial \theta(t,z)}{\partial t} + u \frac{\partial \theta(t,z)}{\partial z}\right) +\dot{q}+h \left(\theta(t,z)-T(t,a)\right)=0\,.
\label{tunnel equation}
\end{equation}
Here $M=\rho_a a/2$ is the mass of air per unit area of tunnel wall and $C_p$
is the specific heat of air in the tunnel, assumed to be a constant for the
temperature ranges under consideration.

The first term represents the change in air temperature with time due to the
thermal mass of the air in the tunnel, heat flux through the tunnel by
convection at an average velocity $u$.  The second term accounts for a general
heat flux $\dot{q}$. This term accounts for the additional flux due to the
short-term fluctuations in air temperature and heat transfer coefficient. Any
other continuous heat sources along the length of the tunnel can also be
accounted for here.  The third term accounts for heat flux through the tunnel
wall due to the temperature difference between the tunnel air, $\theta(t,z)$
and the wall surface temperature along the tunnel, $T(a,t;z)$. This heat flux
is assumed to be purely radial (i.e., no heat flux through the surrounding
soil along the tunnel $z$ axis). As a result, this term can be expressed as
the Bessel formulation of the heat conduction given previously. Hence the
tunnel wall flux can be considered as a complex term multiplied by the tunnel
air temperature $(\dot{q}_{\mathrm{r}}+i\,\dot{q}_{\mathrm{i}}) \theta(t,z)$, where $\dot{q}_{\mathrm{r}}$ and $\dot{q}_{\mathrm{i}}$ are given by
\eqref{periodic flux preliminary} and the driving air temperature term
$\theta(t,z)$ is now a function of the axial distance down the tunnel, so
replaces $f(t)$.

With the above assumptions, the tunnel air temperature can be separated into a
function of the tunnel axis and a time oscillating component as
\begin{equation}
\theta(t,z) = \exp(i \omega t) Z(z)\,.
\end{equation}
The solution to Eq.~\eqref{tunnel equation} is then
\begin{equation}
\theta(t,z) = \frac{b}{a} \left(1- \exp(-a z)\right)+D\exp(i \omega t - a z),
\label{eq:thetatz}
\end{equation}
where
\begin{equation}
a = \frac{1}{u} \left(\frac{A}{M C_p}+i\, \left(\omega+ \frac{B}{M
  C_p}\right)\right), \qquad
b = \frac{\dot{q}}{M C_p u}\,,
\end{equation}
and we applied boundary condition~\eqref{eq:BCz0} at $z = 0$.

Extracting the real part of~\eqref{eq:thetatz} gives the explicit solution for
air temperature within the tunnel as a function of time and axial distance,
\begin{multline}
\theta(t,z) = 
\frac{\dot{q}}{M C_p (E^2+F^2)}\\ \times\left[E - E \exp \left(-\frac{zE}{u}\right) \cos \left(\frac{z F}{V}\right) + F \exp \left(-\frac{z E}{V}\right) \sin \left(\frac{z F}{V}\right)\right]
\\ + D \exp \left(-\frac{z E}{V}\right) \cos \left(\omega t - \frac{z F }{V}\right)
\label{theta}
\end{multline}
where
\begin{equation}
E = \frac{A}{M C_p}\,, \qquad F=\omega+\frac{B}{M C_p}\,.
\end{equation}

The solution given by \eqref{theta} has two distinct components. The first
comes from the inhomogenous term in the governing equation and hence is
proportional to $\dot{q}$. This gives the effect of a continual heat source or
sink along the tunnel which is independent of either the air or tunnel wall
temperature. Such loads can include heat from lighting or other equipment
along the tunnel or the additional heat flux due to the correlation between
short-term fluctuations in air temperature and heat transfer coefficient, as
given in Section~\ref{sec:notindep}.  The second component shows that
fluctuations in the inlet air temperature decay exponentially along the tunnel
axis as $\exp(-{z A}/{M C_p u})$.

\begin{figure}[h!]
\begin{center}
\includegraphics[scale=0.5]{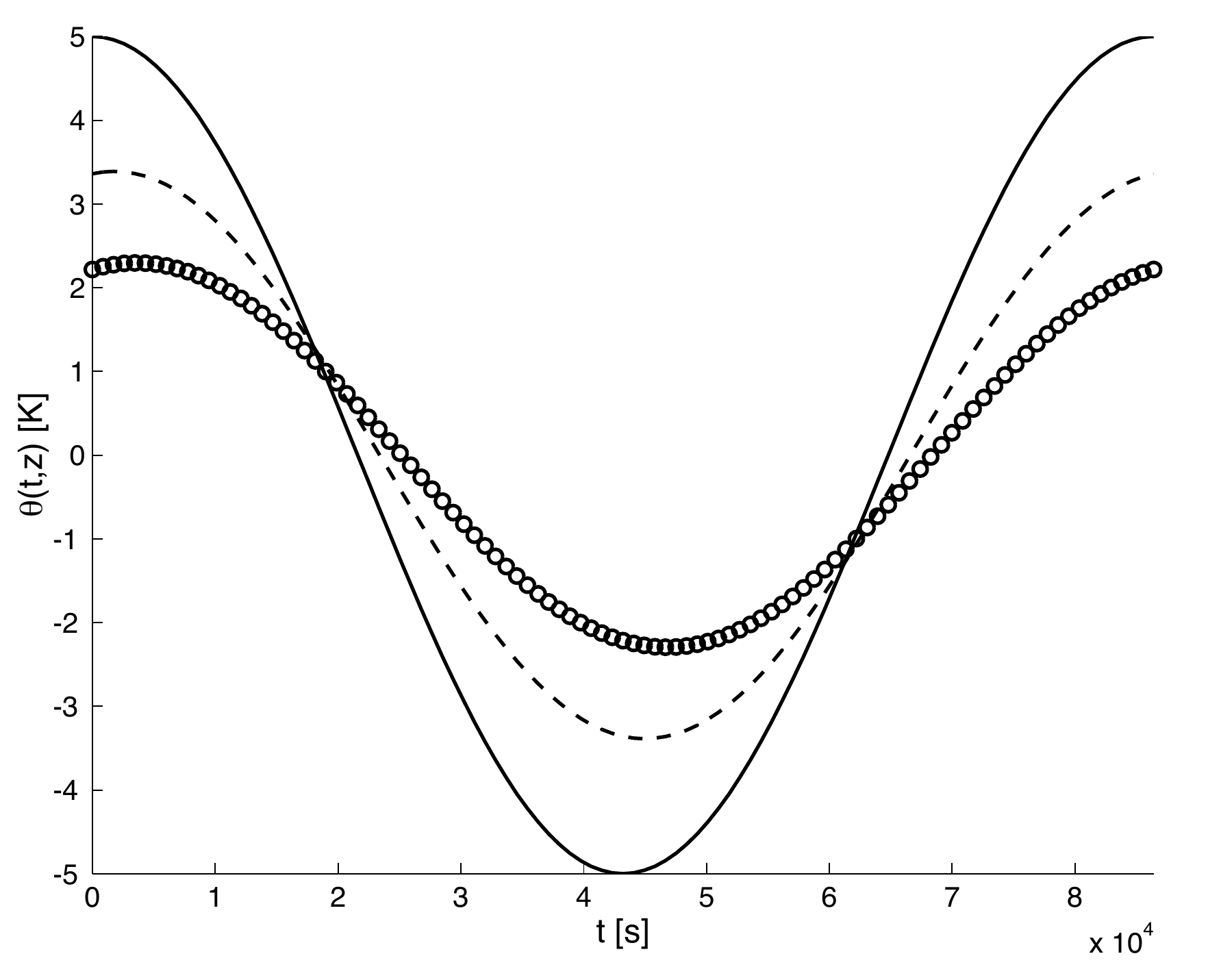}
\caption{Daily temperature fluctuations at the beginning (--), middle (-~-)
  and end (o) of a tunnel.}
\label{graph_theta_phase_daily}
\end{center}
\end{figure}

Figure~\ref{graph_theta_phase_daily} shows inlet air temperature fluctuations
of amplitude 5 degrees Celsius at the beginning, middle and end of a 1000 m
long tunnel for daily temperature variations. The temperature fluctuations are
damped as they propagate along the tunnel. There is also a phase shift in the
fluctuations along the tunnel.

For the parameters involved in this particular application the lower frequency
yearly oscillations are almost unaffected by the thermal mass of the
soil. This means that, once the initial transients die away, seasonal
oscillations in ambient air temperature do not affect the peak air temperature
in the tunnel. The majority of the damping is due to daily oscillations. This
was suggested by the results presented in Section \ref{sec:conduction2} (see
Figure~\ref{graph_picc_transient} where the tunnel surface temperature closely
approached the air temperature and hence greatly reduced heat flux over the
wall).

\begin{figure}[h!]
\begin{center}
\includegraphics[scale=0.5]{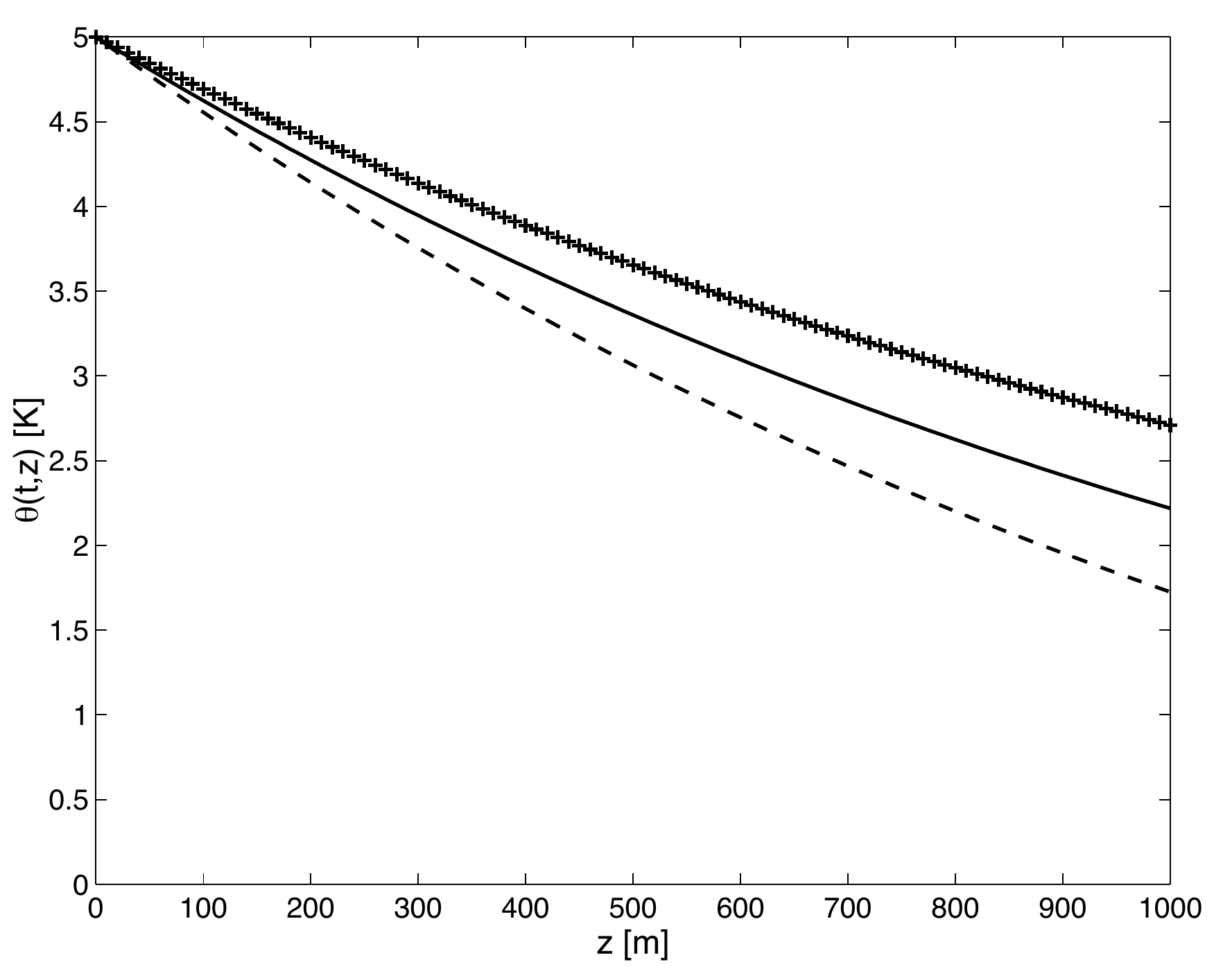}
\caption{Decay of peak daily temperature along a tunnel with varying tunnel
  heat load.}
\label{graph_theta_decay}
\end{center}
\end{figure}

Figure~\ref{graph_theta_decay} shows the decay of the peak daily temperature
along the length of the tunnel. By the end of the tunnel (1000 m) the
amplitude of the fluctuations is approximately half this value.
The effect of the inhomogeneous term $\dot{q}$ is also shown in
Figure~\ref{graph_theta_decay}. The three different profiles are for
$\dot{q}=$ 0, -10 (-~-) and +10 (+) W/m$^2$/m (Watts per square meter of tunnel surface per meter of tunnel length). Depending on the sign of the flux
term, the air temperature along the tunnel is shifted either up or down, as expected.  

\section{Summary} 
\label{sec:summary}

This paper has presented analytic solutions to a number of air flow, heat
convection and heat condition problems. These include the following:
\begin{itemize}
\item A model of the turbulent air flow profile through both an open circular
tunnel and in the annulus between a stationary circular tunnel with a moving
concentric core. This is used to estimate the air flow profile and hence the
convective heat transfer coefficient for an open tunnel and between the tunnel
wall and a moving train.
\item The evolution of the temperature of the soil surround an underground
tunnel in response to an instantaneous change in air temperature within the
tunnel. This is used to estimate the period of time it takes for the soil
around a new tunnel to reach a steady operating temperature.
\item The limit cycle thermal response of the soil surrounding a tunnel to an
oscillating air temperature within the tunnel. This is used to quantify the
effect of the thermal mass of the soil on temperatures within the tunnel.
\item The effect of a correlation between short-term fluctuations in air
temperature and heat transfer coefficient, both the result of passing trains,
is to increase the mean soil temperature.
\item Finally, a model is derived that accounts for the propagation of
temperature fluctuations along the axis a tunnel of finite length and through
the surrounding soil (perpendicular to the tunnel axis).
\end{itemize}

The analytical model for the coupled evolution of air temperature within and
soil temperature around an underground rail tunnel can be applied to a simple
rail tunnel as described, or to other engineering applications where the
effect of periodic temperature variations influence an effectively infinite
solid. These include modelling the heat dissipation from a cable tunnel (used
for underground high voltage power distribution) and the response of the soil
temperature to ground coupled heat pumps.  The more detailed treatment of the
thermal mass surrounding a tunnel would allow existing models, such as that
developed by \citet{Ampofo}, to quantify the effects potential benefits of
flushing the London Underground with cool night air, for example.

Further work on this subject could be undertaken to apply the underlying
analytical approach to more complicated boundary conditions. While some
tunnels are lined with concrete, which is thermally similar to the surrounding
soil, many tunnels are lined with cast iron segments, which are not. An
obvious progression of the work presented in this paper is to make an
allowance for tunnel linings. Further, tunnels are often within the thermal
boundary layers of adjacent tunnels or the ground surface. As a result the
assumption that the tunnel is surrounded by an infinite amount of soil with no
additional heat loads is not valid.

\clearpage

\appendix

\section{Physical Parameters}
\label{apxA}

\FloatBarrier

\begin{table}
\caption{Physical constants from \citet{kreith}}
\label{table:constants}
\begin{tabular}{p{5cm}p{0.7cm}p{2cm}p{1.4cm}p{5cm}}
\hline
Density (air) & $\rho_a$ & 1.16 & kg/m$^3$ & \\	
Absolute Viscosity (air) & $\mu$ & 1.82$\times 10^{-5}$ & N.s/m$^2$ & \\
Dynamic Viscosity (air) & $\nu$ & 1.57$ \times 10^{-5}$ & m$^2$/s &
$\nu=\mu/\rho_a$\\
Thermal Conductivity (air) & K$_a$ & 2.51$ \times 10^{-2}$ & W/m.K & \\
Specific Heat (air) & C$_{pa}$ & 1012 & J/kg.K & \\
Density (soil) & $\rho_s$ & 1500 & kg/m$^3$ & \\
Specific Heat (soil) & C$_{ps}$ & 1842 & J/kg.K & \\
Theramal Conductivity (soil) & K$_s$ & 0.35 & W/m.K & \\
Thermal Diffusivity (soil) & $\kappa$ & 1.27 $ \times 10^{-7}$ & m$^2$/s & \\
\hline
\end{tabular}
\end{table}

\begin{table}
\caption{Piccadilly Line Parameters from \citet{UC}}
\label{table:piccadilly}
\begin{tabular}{p{5cm}p{0.7cm}p{2cm}p{1.4cm}p{5cm}}
\hline
Tunnel Radius & $a$ & 1.70 & m & \\
Tunnel CSA & $A_t$ & 9.07 & m$^2$ & $At=\pi a^2$\\
Train CSA & $A_v$	 & 6.00 & m$^2$ & \\
Train Radius (equivalent) & $b$ & 1.38 & m & $b=(A_v/ \pi)^{1/2}$\\
Annulus Area & $A_a$ & 3.07 & m$^2$ &$A_a=A_t-A_v$\\
Wall Roughness & $k_o$ & 0.01 & m & typical\\
Train Surface Roughness & $k_i$ & 0.01 & m & typical\\
Train Speed & $V$ & 14.0 & m/s & typical\\
Average Tunnel Air Velocity & $U_t$ & 10.0 & m/s & assumed\\
Tunnel Air Flux & $Q$ & 90.7 & m$^3$/s & $Q=U_t\times A_t$\\
Average Annulus Air Velocity & $U_a$ & 29.5 & m/s & $U_a=Q/A_a$\\
Train Headway & $hwy$ & 120 & s & time between trains\\
Train Length & $L$ & 183 & m & \\
Train Passing Time & $t_{pass}$ & 13.1 & s & $t_{pass}=L/V$\\
Prop. of time Train Present & $\sigma$ & 0.11 & - & $\sigma=t_{pass}/hwy$\\
Angular Freq. (short-term) & $\omega_\varepsilon$ & 5.23 $\times 10^{-2}$ & rad/s & $\omega_\varepsilon = 2 \pi / hwy$\\
\hline
\end{tabular}
\end{table}

\begin{table}
\caption{Open Tunnel Flow}
\label{table:tunnel}
\begin{tabular}{p{5cm}p{0.7cm}p{2cm}p{1.4cm}p{5cm}}
\hline
Reynolds Number & $Re$ & 1.08 $\times 10^6$ & - & $Re=U_t a/ \nu$ \\
Friction Velocity & $v_*$ & 0.615 & m/s & $v_*=Q/(2 \pi
a^2(2.375+1.25\log(1/k_o))$\\
Flow Type & $F_t$ & 392 & - & $F_t=v_*k_o/ \nu$ \\
Wall Shear Stress & $\tau_w$ & 0.439 & N/m$^2$ & $\tau_w=\rho_a v_*^2$\\
Tunnel Heat Transfer Coeff. & $h_t$ & 44.4 & W/m$^2$.K & $h_t =\tau_w
C_{pa}/U_t$\\
\hline
\end{tabular}
\end{table}

\begin{table}
\caption{Train / Tunnel Annulus Flow}
\label{table:annulus}
\begin{tabular}{p{5cm}p{0.7cm}p{2cm}p{1.4cm}p{5cm}}
\hline
Reynolds Number & $Re$ & 1.20 $\times 10^6$ & - & $Re=2 U_a (a-b)/ \nu$\\
Inner flow integral & $A_i$ & 2.00 & - & $A_i=d_i (6b+3.625d_i+(2.5
b+1.25d_i)(\log(1/k_i)))$\\
Outer flow integral & $A_o$ & 6.60 & - &
$A_o=d_o(6a+3.625d_o+(2.5a-1.25d_o)(\log(1/k_o)))$\\
Inner Friction Velocity & $v_{*i}$ & 0.958 & m/s & $v_{*i} =
v_{*o}(d_i/d_o)^{1/2}$\\
Inner Boundary Layer & $d_i$ & 8.01 $\times 10^{-2}$ & m & Solved
Numerically\\
Outer Friction Velocity & $v_{*o}$ & 1.66 & m/s & $v_{*o} = (Q-2 \pi
v_{*i}A_i-\pi V v_{*i} (d_i^2+2bd_i))/(2 \pi A_o)$\\
Outer Boundary Layer & $d_o$ & 0.24 & m & $d_o = a-b-d_i$\\
Flow Type (wall) & $F_{tw}$ & 1057 & - & $F_{tw}=v_{*o} k_o/ \nu$\\
Flow Type (train) & $F_{tt}$ & 611 & - & $F_{tt}=v_{*i} k_i/ \nu$\\
Wall Shear Stress & $\tau_w$ & 3.19 & N/m$^2$ & $\tau_w=\rho_a v_{*o}^2$\\
Train Shear Stress & $\tau_t$ & 1.06 & N/m$^2$ & $\tau_t=\rho_a v_{*i}^2$\\
Annulus Heat Transfer Coeff. & $h_v$ & 110 & W/m$^2$.K & $h_w =\tau_w
C_{pa}/U_a$\\
\hline
\end{tabular}
\end{table}

\begin{table}
\caption{Effective Heat Transfer Coefficient}
\label{table:heattransfer}
\begin{tabular}{p{5cm}p{0.7cm}p{2cm}p{1.4cm}p{5cm}}
\hline
Open Tunnel Coefficient & $H_t$ & 127 & m$^{-1}$ & $H_t=h_t/K_s$\\
Non-dim decay time & $t_d$ & 0.20 & - & $t_d=\kappa t / a^2$\\
Decay Time (open tunnel) & $t$ & 4.56 $\times 10^6$ & s & $t=t_d a^2/ \kappa$\\
Annulus Coefficient & $H_v$ & 314 & m$^{-1}$ & $H_v=h_v/K_s$\\
Non-dim decay time & $t_d$ & 0.02 & - & $t_d=\kappa t/a^2$\\
Decay Time (annulus) & $t$ & 4.56 $\times 10^5$ & s & $t=t_d a^2/ \kappa$\\
\hline
\end{tabular}
\end{table}

\begin{table}
\caption{Decay of Temperature Fluctuations with Radial Distance}
\label{table:tempdecay}
\begin{tabular}{p{5cm}p{0.7cm}p{2cm}p{1.4cm}p{5cm}}
\hline
Non-dim. decay distance & $d_d$ & 2.2 & - & from Figure 3.5\\
Period (yearly oscillations) & $p_y$ & 3.15 $\times 10^7$ & s & \\
Angular Freq. (yearly) & $\omega_y$ & 1.99 $\times 10^{-7}$ & rad/s & $\omega_y=2 \pi/p_y$\\
Therm. bndry layer (yearly) & $L_y$ & 1.8 & m & $L_y=d_d/(\omega_y/ \kappa)^{1/2}$\\
Period (daily oscillations) & $p_d$ & 8.64 $\times 10^4$ & s & \\
Angular Freq. (daily) & $\omega_d$ & 7.27 $\times 10^{-5}$ & rad/s & $\omega_d=2 \pi/ p_d$\\
Therm. bndry layer (daily) & $L_d$ & 0.1 & m & $L_d=d_d/(\omega_d / \kappa)^{1/2}$\\
\hline
\end{tabular}
\end{table}

\clearpage



\end{document}